\documentclass[12pt,preprint]{aastex}

\usepackage{xcolor} 
\usepackage{ulem}   

\slugcomment{Revision: Last updated \today}

\shorttitle{Dwarf galaxies in NGC 3585 Group}
\shortauthors{Park et al.}

\begin{document}

\title{Dwarf Galaxy Discoveries from the KMTNet Supernova Program{*} II. The NGC 3585 Group and Its Dynamical State}

\author{ Hong Soo Park\altaffilmark{1}, 
Dae-Sik Moon\altaffilmark{2},
Dennis Zaritsky\altaffilmark{3}, 
Sang Chul Kim\altaffilmark{1,4},
Youngdae Lee\altaffilmark{1},
Sang-Mok Cha\altaffilmark{1,5}, and
Yongseok Lee\altaffilmark{1,5}
}

\email{hspark@kasi.re.kr
}

\altaffiltext{1}{Korea Astronomy and Space Science Institute, 776, Daedeokdae-ro, Yuseong-gu, Daejeon 34055, Korea}
\altaffiltext{2}{Dept. of Astronomy and Astrophysics, University of Toronto, 50 St. George Street, Toronto, ON M5S 3H4, Canada}
\altaffiltext{3}{Steward Observatory, University of Arizona, 933 North Cherry Avenue, Tucson, AZ 85719, USA}
\altaffiltext{4}{Korea University of Science and Technology (UST), Daejeon 34113, Korea}
\altaffiltext{5}{School of Space Research, Kyung Hee University, Yongin, Kyeonggi 17104, Korea}

\altaffiltext{*}{Based on data collected at the KMTNet Telescopes}
 
\begin{abstract}

We present our discovery and analysis of 
dwarf galaxies in the 
NGC 3585 galaxy group by the KMTNet Supernova Program.
Using deep stack images reaching $\simeq$ 28 mag arcsec$^{-2}$ in $BVI$,
we discovered 46 
dwarf galaxy candidates 
distributed in a
7 square degree field. 
The dwarf galaxy candidates exhibit 
 central surface brightness as faint as $\mu_{0,V} = 26.2$  mag arcsec$^{-2}$,
with 
  effective radii larger than 150 pc and 
  total absolute magnitudes brighter than $M_V \approx -10$ mag, if
  at the distance of NGC 3585.
The dwarf galaxy surface number density decreases with projected distance from NGC 3585.
We estimate the background contamination to be about 20\% 
\textcolor{black}{based both on the number density profile and on diffuse galaxy counts in a control field.}
The dwarf galaxy colors and S{\'e}rsic structural parameters are consistent with 
those found for other dwarf galaxies. 
Unusually, there is no indication 
  of a change of color or brightness in the dwarf galaxy candidates
with projected distance from the group center. 
Approximately 20\% of them contain an unresolved nucleus. 
The nucleated fraction 
is larger for brighter (and redder) galaxies
 but is independent of distance from the group center. We identify
four ultra-diffuse galaxy candidates, all 
near 
the group center.
We interpret these spatial properties as  suggesting that the NGC 3585 group \textcolor{black}{might be} dynamically younger than the typical group.
%
The galaxy luminosity function 
of the NGC 3585 group has 
a faint-end slope of $\alpha\approx -1.39$, 
\textcolor{black}{
which is roughly consistent with the slopes found for other nearby groups.
The possible dependence of the slope on global group properties is still unclear,
and continues to motivate our homogeneous deep survey of dozens of nearby groups.}
\end{abstract}


\section{Introduction}

%
Dwarf galaxies in galaxy groups are
  excellent tools for exploring galaxy formation and evolution in the critical
  intermediate-density environments
  \citep{ben02, coo10, beh10, wet12, pre12} and
  the detailed predictions of cosmological models \citep{kly99,moo99,tre02}.
Dwarf galaxy studies have 
  traditionally focused on the Local Group (LG). 
  However, technological advances have enabled wide field surveys for low surface brightness dwarf galaxies at larger distances \citep{mcc12,bec15,drl15}.
For example, 
 access to wide-field facilities ($\gtrsim$ 1 $\square^\circ$ such as with the Canada-France-Hawaii Telescope and the Blanco 4-m Telescope) allows investigators to 
 measure the luminosity function
 in nearby ($D\lesssim$ 10 Mpc) groups to 
$M_V\sim -10$ mag 
 (e.g. M101 \citep{mer14, mul17b, ben17},
   M96 \citep{mul18},
   M81 \citep{chi09},
   M83 \citep{mul15},
   NGC 5128 \citep{crn16, mul17a}). 
 %
 
 With such data, we can begin to compare
 key properties of dwarf galaxies, 
  such as the range of structural parameters, 
 the incidence of nucleated or ultra-diffuse galaxies, the luminosity function, and the spatial distribution,
 among the set of nearby groups   \citep{zab00,chi09, tre09, mer14, van15, crn16, mul15, mul17b, mul18, par17} and to the properties of dwarfs found in the field and cluster environment   \citep{you14,jan16,lee18} to search for 
the effect of environmental influences.
However, despite all of these recent developments, 
it is still difficult to obtain a consistent overview 
because 
 comparisons 
  are hindered by
  systematic differences in data quality, selection criteria, and technical methodology among the studies. 

\textcolor{black}{
Several ongoing surveys are attempting to address this topic.
Recently, \citet{geh17} introduced the Satellites Around Galactic Analogs (SAGA) Survey, which aims
to investigate the distribution of dwarf galaxies around 100 Milky Way (MW)-like systems at $20-40$ Mpc
down to a satellite luminosity of $M_r \approx -12$ mag. 
They 
reported the results from the study of eight MW-analog hosts, 
  including spectroscopically confirmed complete dwarf galaxy luminosity functions
  using photometry based on the SDSS $gri$-bands.    
In another survey, \citet{gre18} presented 800 faint dwarf galaxies 
  identified in 200 deg$^{2}$ of the Hyper Suprime-Cam Subaru Strategic Program (HSC-SSP),
  a program with a plan of eventually covering 1400 deg$^{2}$ with $grizy$-bands during 300 nights \citep{aih18}.
Previously, \citet{mer14} reported the result of the faint dwarf galaxies around the M101 galaxy
 using the observations taken with the Dragonfly Telephoto Array.
Using the results from other Dragonfly observations, there have been several studies 
on the properties of nearby groups
\citep{dan17,coh18}. Other relevant studies will be discussed in more detail farther below.
Each study, however, has its own strengths and weaknesses. For example, while spectroscopic information is valuable in that it helps confirm membership, it 
also results in a brighter limit on the galaxies that are considered as satellites (e.g., SAGA). Other times, high sensitivity to 
low surface brightness objects comes at the expense of angular resolution, which can limit the study of the galaxies' structural properties
(e.g., Dragonfly). And yet other times, the limitation is something as simple as limited hemispheric coverage. While the last may seem to be minor, it can be important in some contexts. In our area of interest, going as far down the luminosity function as possible in groups, there are a limited number of suitable objects in the sky and accessing both hemispheres is helpful.
%
}

%

We are conducting a systematic study of dwarf galaxies
in nearby galaxy groups 
 using data from the KMTNet (Korea Microlensing Telescope Network) Supernova Program \cite[KSP;][]{moo16}.
KSP is 
a program to search for supernovae and
   optical transients \citep{he16,ant17,bro18,lee19,afs19}
 using three 1.6-m telescopes located at
 the Cerro Tololo Inter-American Observatory (CTIO, Chile),
 the South African Astronomical Observatory (SAAO, South Africa),
 and the Siding Spring Observatory (SSO, Australia).
When several hundred KSP images in each field 
%
  are stacked, 
  the program 
  provides deep $BVI$ images reaching a sensitivity of about $28$ mag arcsec$^{-2}$
  within a 1.5 acrsec radius aperture.
%
\textcolor{black}{
The KSP dwarf galaxy survey investigates the properties of group dwarf galaxies down to total absolute magnitude, $M_V\sim -10$ mag, using multi-band ($BVI$) imaging. 
Our survey area covers non-bulge seasons, $22^h< {\rm R.A.} <24^h$ and $0^h< {\rm R.A.} <14^h$ for 
  ${\rm Decl.} <0\arcdeg$. 
Our target fields 
  normally have a main galaxy that is brighter than $M_V\sim -20$ mag and relatively nearby, $D<20$ Mpc.
   Additionally, we select the groups with a velocity dispersion, $\sigma_{group}$, that is 
     $>60$ km s$^{-1}$ or a total mass that satisfies $\log({\rm M/M_\sun})>12.0$ \citep{mak11}.
     We are observing more than 30 such galaxy groups up to at least 2023, when the 
     commissioning season for the KMTNet Phase-2 is scheduled to close.
     %
Normally, for each targeted group we observe two fields, 
     which are observed through two seasons and
     overlap so that the main galaxy is included in one chip (see Section \ref{data} in detail). 
     This results in coverage where the maximum projected radial distance from the group main galaxy is roughly $>$0.5 Mpc.
%
Our aims are to 1) identify new (and previously known) dwarf galaxies, 2) identify any cases with unusual structures
  (e.g. nucleated dwarf galaxies, UDGs, star clusters), and 3) uncover how group properties relate to the properties of the
  satellite population, such as the spatial distribution and luminosity function.}
In our first paper describing results from these data \citep{par17},
we reported our discovery and analysis of $\sim$ 30
dwarf galaxy candidates toward NGC 2784.
%
The eventual, complete analysis of the KSP data will
result in homogeneous results that can be compared
for \textcolor{black}{$>30$} nearby groups.

In this paper, we report on the discovery and analysis of dwarf galaxy candidates 
  in the NGC 3585 field 
  using stacked KSP images.
The NGC 3585 group is known to include NGC 3585 itself, which is an E6 galaxy and the brightest galaxy in the group, and at least eight other galaxies with consistent radial velocities \citep{mak11, tul13}.
 We adopt a distance to this group of 20.4 Mpc that was estimated using the surface brightness fluctuation (SBF) method \citep{tul13}.
  In Section 2, we present the observations and data reduction.
  In Section 3, we describe our search and the photometric results for the dwarf galaxy candidates.
  In Section 4, we discuss the properties of dwarf galaxy candidates detected in the NGC 3585 group and compare them to those of dwarfs in other groups.

\section{Observations and Data Reduction}\label{data}
 
\begin{figure}
\epsscale{1.0}
\plotone{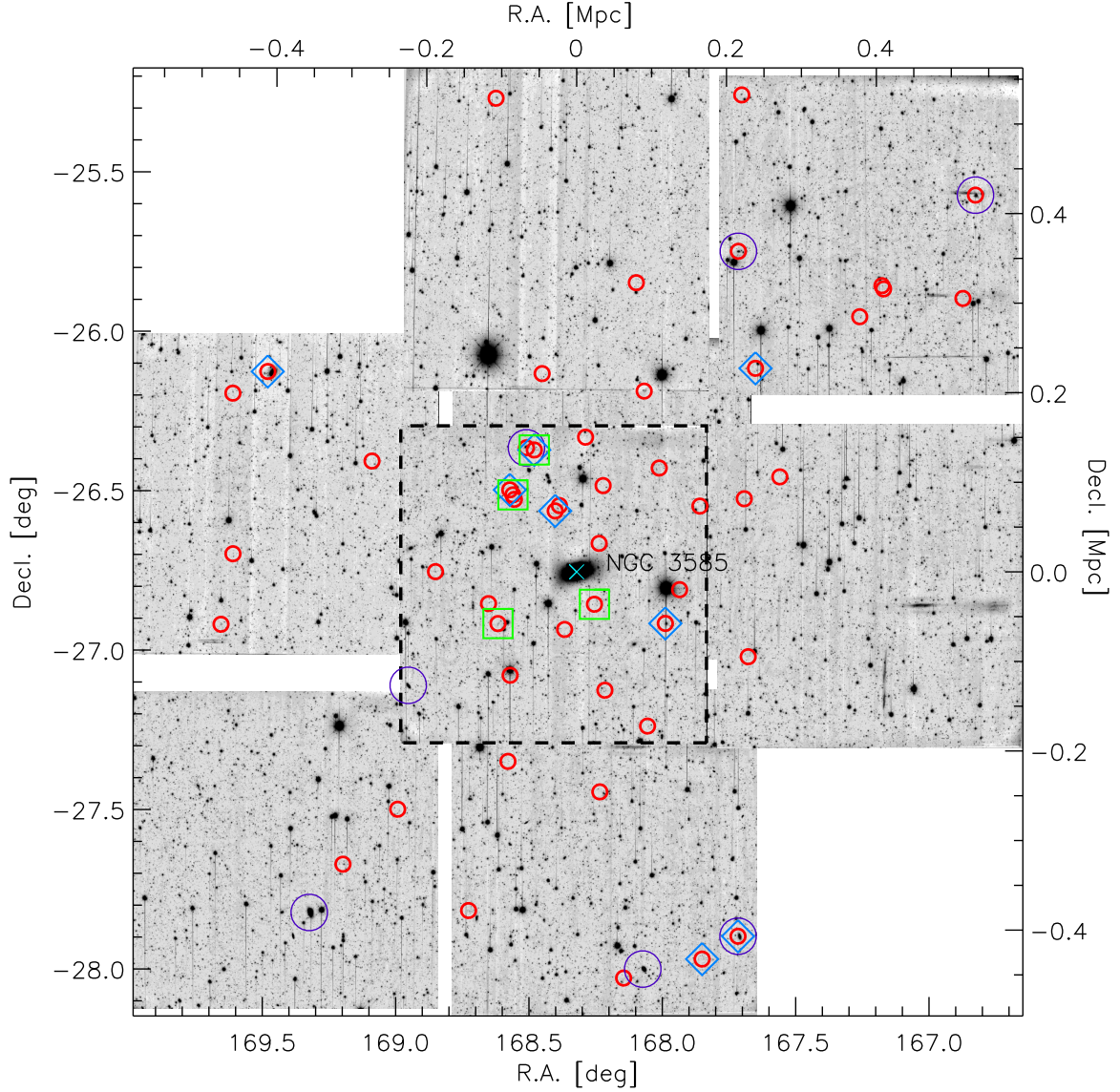}
\caption{ 
 \textcolor{black}{$I$-band stacked image of the entire NGC 3585 field observed by KMTNet.
 The red open circles are our dwarf galaxy candidates. 
The large dashed square represents one chip of the N3585-1 field
 with approximately $1^\circ \times 1^\circ$ (0.36 Mpc $\times$ 0.36 Mpc) area.
The large purple open circles indicate the galaxies with radial velocities 
 similar to that of NGC 3585.
The blue diamonds and green boxes are the nucleated dwarf galaxy candidates and the UDG candidates classified in this study, respectively. 
The cross indicates the center of the NGC 3585 galaxy.
 } 
\label{fig-radec}}
\end{figure}

We use 
  $BVI$ observations of two fields including  NGC 3585 
 from the KMTNet Supernova Program
 using the three KMTNet telescopes \citep{kim16} from January 2016 to June 2017.
The wide-field CCD camera installed in each telescope 
covers 
4 $\square^\circ$ (2$^\circ \times 2^\circ$) with 0.4$\arcsec$ pixel$^{-1}$ scale.
\textcolor{black}{
The detector is a mosaic of four chips and
each chip has 9K $\times$ 9K array with 
with a small gap between chips. Each chip corresponds to a
 $1^\circ \times 1^\circ$ field-of-view.}
$BVI$ filters were used and the exposure time in each image was fixed to be 60 seconds.
In Figure \ref{fig-radec} we show \textcolor{black}{the $I$-band stacked image for} 
  the two fields observed toward NGC~3585, which
 overlap in an area of approximately 1 deg $\times$ 1 deg 
 \textcolor{black}{(0.36 Mpc $\times$ 0.36 Mpc at the distance of the NGC 3585),} 
so the net size of the field of view is 7 square degrees. In addition, we also show in 
Figure \ref{fig-radec} the locations of our newly discovered dwarf galaxy candidates. One region,
N3585-1 (\textcolor{black}{included as the dashed square} in Figure \ref{fig-radec}), 
  was observed approximately six hundred times in each filter, 
  while the other region, N3585-2 (\textcolor{black}{four images toward lower-left}), was observed about half as many times.
As a result, the former images are approximately 0.3 mag arcsec$^{-2}$ deeper than the latter in $I$-band surface brightness.

Following the standard pre-processing procedure, which includes
a crosstalk correction, overscan subtraction, and flat fielding,
we obtain an astrometric solution using the $SCAMP$ program \citep{ber06}.
We then stack the images that have better than 2$\arcsec$ seeing using the $SWARP$ program \citep{ber02}
to produce the images that we analyze and present here
(see Section 2 of \citet{par17} for more details).
The seeing in the resultant stacked images is better than 1.5 arcsec in all bands.
Details of the observations are listed in Table \ref{tab-obslog}.

We obtain photometric solutions for our stack images
using standard stars in the AAVSO Photometric All-Sky Survey (APASS) catalog \citep{apass}.
As in Park et al. (2017; see their Section 2 and Figure 2 for details), 
 we make a color correction
 in transforming instrumental $B$-band magnitudes to standard values:
 $B = b + c(B-V) + zero$.  
 Here, we use 0.27 for the \textcolor{black}{approximate} color coefficient ($c$), 
 which we obtained from \textcolor{black}{about 150 stars per chip} 
 (the value of $c$ is almost identical to what we obtained for NGC 2784 fields - see Park et al. 2017).
 On the other hand, we apply only zero-point offsets for the transformation of the $V$ and $I$-bands.
The zero-points for each band are 
$28.2 \pm 0.03$, $28.2 \pm 0.03$, and $28.1 \pm 0.05$ mag for $B$, $V$, and $I$-bands, respectively.
\textcolor{black}{
We list the values we use for the standardization of the instrumental magnitudes in each chip in Table \ref{tab-stdmag}.
}
We note that for the $I$-band calibration, we convert the Sloan $r$ and $i$ magnitudes in the APASS catalog to $I$-band magnitudes using the equation, $I=r-1.2444(r-i)-0.3820$ from \citet{lup05}.   
We then correct our galaxy photometry for  Galactic extinction in this field using $A_B=0.231$, $A_V=0.175$, and 
$A_I=0.096$ \citep{sch11}\footnote{To clarify some confusing notation, we only apply extinction corrections to total magnitudes and colors. If such corrections have been applied, then we signify that with a subscript 0, as in $(B-V)_0$. If no subscript is included, then the corrections have not been applied. Second, for historical consistency we also use the subscript 0 to refer to central values, as in the central surface brightness, $\mu_0$. The subscript here does not imply that an extinction correction was applied. Third, we also use the subscript zero on a scaling radius, $r_0$, that is different from the effective radius, $r_e$. The relationship between the scaling radius and $r_e$ depends on the S\'ersic index, $n$.}.


\section{New Dwarf Galaxy Candidates and Basic Properties}\label{result}  

\subsection{Search for Dwarf Galaxy Candidates}\label{dwsearch}

\textcolor{black}{  
 To find dwarf galaxy candidates in the two NGC 3585 fields,
we visually 
select dwarf galaxy candidates by identifying diffuse sources \citep[for details]{par17}.
For dwarf galaxy candidates in the NGC 3585 group, the $\sim$ 20 Mpc distance, or $(m-M)\sim 31.5$ mag,
means that the stellar populations will present as unresolved diffuse features 
 because even the stars on the tip of the red giant branch (TRGB), one of the brightest stellar populations, can not be resolved.
 In our stacked image, which has a point source magnitude limit of about 24 mag,
 the TRGB stars with $M_I=-4.0$ mag \citep{lee93} have an apparent magnitude of about 27.5 mag.
We use primarily the $I$-band image to identify galaxies
  because it is the deepest among the three bands for a given exposure time
  and because most dwarf galaxies have a red color, $(V-I)>0.5$ (see Section \ref{cmdstructure}).
We use the $B$ and $V$-band images, as well as RGB color images made from the three bands, to rule out an artifacts.
}

\textcolor{black}{ 
To enhance faint surface brightness features,
we set the dynamical scale range on a display screen (e.g. $DS9$) such that
 the lowest and the highest intensities are approximately $-3\sigma$ and $10\sigma$ from the mean, where $\sigma$ refers to the scatter of the sky background, respectively.
  We select any diffuse sources that are approximately larger than $10\arcsec$ as dwarf galaxy candidates,
      and currently reject sources with special features (e.g. spiral, bulge, etc) except that those 
   with a point source at the center are classified as nucleated dwarf galaxy candidates (see Section \ref{nucleus}).
To minimize 
biases in the detection and incompleteness, 
  three authors (H.S.P., S.C.K., and Y.L.) searched the images independently and then compared the results. 
For candidates selected by only one or two classifiers, 
all three classifiers reviewed together
the object and decided whether to retain or reject the candidate
basically based on the above criteria.
The retained candidates are roughly 20 percent of all the candidates finally accepted.
}

For the regions near NGC 3585 itself, 
  we use an image where we have subtracted a model for the galaxy constructed with the  IRAF/$BMODEL$ task.
\textcolor{black}{
The model is constructed using the output from the IRAF/$ELLIPSE$ task,
   utilizing non-linear steps along the semimajor axis and 
   floating ellipticity and position angle,
   with the brightness profile measured using median isophotal fluxes
   after interactively masking bright objects (e.g. foreground stars or background galaxies) and 
   applying two iterations of 3$\sigma$ clipping.
This model subtraction is done only on the 0.25 deg $\times$ 0.25 deg area centered on NGC 3585 galaxy in the N3585-1 field.
From our visual inspection of the subtracted image we conclude that the model fitting is adequate to within 1 arcmin from the center of NGC 3585.
%
}   

Despite the difference in depth between the two observed fields, we recover the same candidates in both fields within the overlapping region. We conclude that our ability to recover dwarfs is not set by the point source limiting magnitude of each field, but rather by other factors such as scattered light from bright stars, image defects, and a spatially variable background. Because none of these limiting factors are affected by the modest difference in exposure time between the two fields, we treat the two fields identically.
\textcolor{black}{
We also searched for dwarf galaxy candidates on the image stacked within the overlapping region,
but did not find any new diffuse sources.} 
In total, we identify 46 such diffuse objects and plot their positions in Figure \ref{fig-radec}. 

We then use the {\it NASA Extragalactic Database (NED)}
  to search for known galaxies in the NGC 3585 KMTNet fields.
We find eight galaxies  with existing radial velocity measurements that are 
compatible with membership in the NGC 3585 group as defined by being within $\pm 3\sigma_v$, 
1220 $< cz < $ 1650 km s$^{-1}$.
 We define this velocity range using the measured group velocity
   dispersion, $\sigma_{\upsilon}=70$ km s$^{-1}$ \citep{mak11} and 
   the radial velocity, $\upsilon=1434$ km s$^{-1}$ ({\it NED}).
Among these eight galaxies, the four that have an absolute magnitude fainter than $-18$ mag
and approximately exponential brightness profiles are dwarf galaxy group members. 
These galaxies are, in retrospect, \textcolor{black}{detected} 
    in our KMTNet images, but do not belong to the 46 dwarf candidates. 
They were not included in our candidate list because they have high surface brightnesses
  and are much more difficult to distinguish from background galaxies. 
\textcolor{black}{
That is, these bright dwarf galaxies have an extended central region (approximately $\mu\lesssim 23$ mag arcsec$^{-2}$) that is saturated on our displayed range, which we generally consider to be a signature of background galaxies.}
As such, there could be additional high surface brightness group members that are neither among the NED galaxies or our cataloged candidates.

\textcolor{black}{
We also examined whether archival images, such as those from the Dark Energy Camera Legacy Survey (DECaLS) and the $HST$,
could be used to confirm our candidates and to search for new diffuse sources. 
Unfortunately, the survey region of the former does not overlap with our NGC 3585 field.
The latter, $HST$, has two overlapping fields (a central region and a region $\sim30$ arcmin away from the NGC 3585 galaxy).
However, none of the dwarf galaxy candidates we identified lie within those fields and 
we did not find any additional diffuse candidates.}

\begin{figure}
\epsscale{1.0} 
\plotone{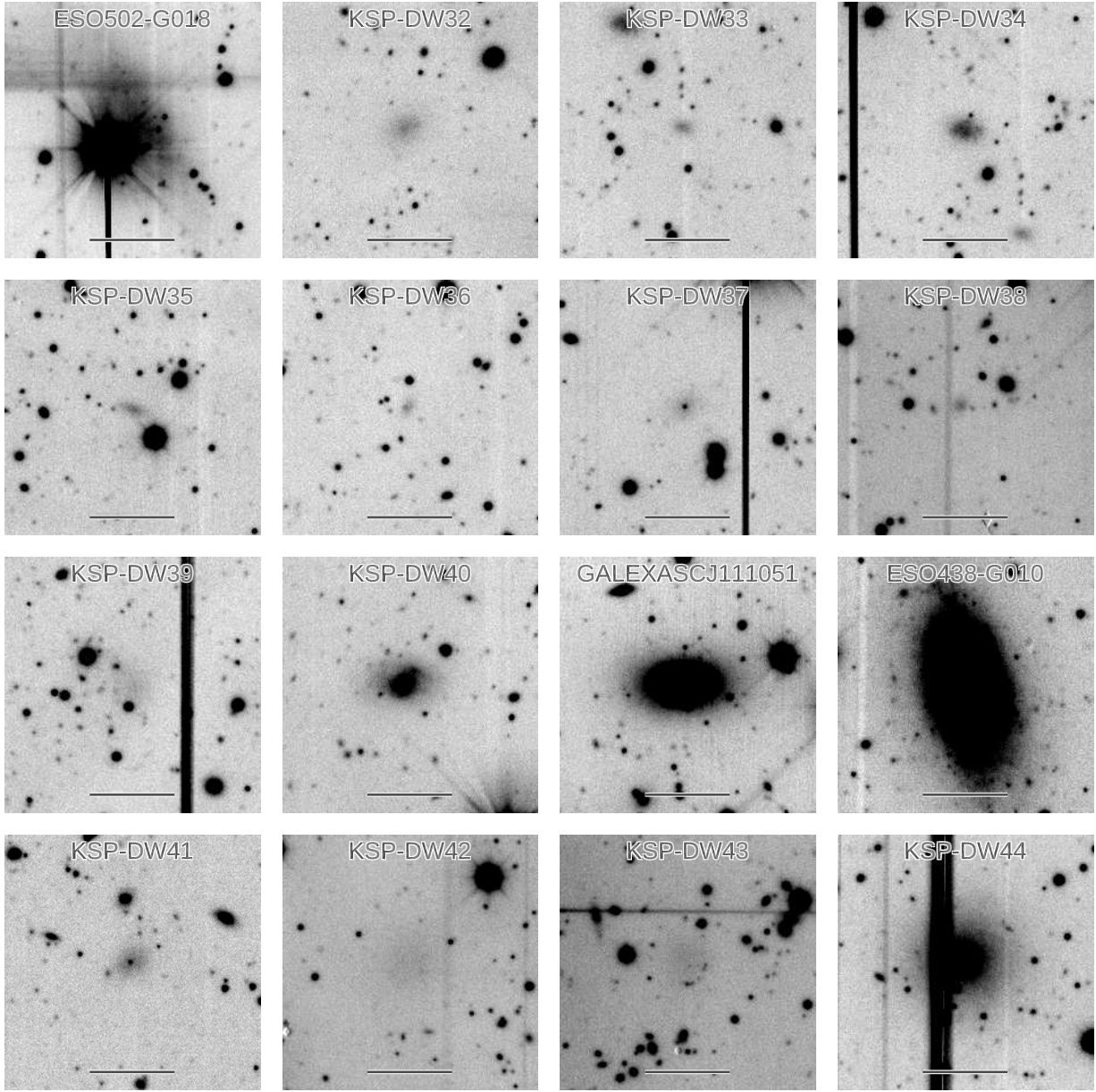}
\caption{ 
$I$-band images of the dwarf galaxy candidates in N3585 fields. 
The scale bar in each panel represents $30\arcsec$ and  
the field-of-view of each image is $\sim 1.5\arcmin \times 1.5\arcmin$.
North is up and east is to the left.
\label{fig-graymapa}}
\end{figure}

We now add the four $NED$ dwarfs to our dwarf galaxy candidate list for the remainder of this study.
We use the brighter four confirmed group galaxies, which are not dwarfs, only when discussing the galaxy luminosity function in Section 4.4.
In Figure \ref{fig-graymapa} we present the $I$-band greyscale images of all the dwarf galaxy candidates
and the additional 4 confirmed dwarfs.
They tend to be diffuse with a variety of sizes and morphologies.
Some show concentrated nucleated emission.
%

We estimate the limiting brightness of our $I$-band images using a completeness test where we create and attempt to recover several hundred artificial galaxies. 
We create the galaxies
  \textcolor{black}{ and add them to our $I$-band images of 
  the deep field (N3585-1-Q2) and the shallow field (N3585-2-Q1),
  including the NGC 3585 galaxy,}
  using the {\it mkobjects} task in the IRAF{\it /artdata} package. 
In our modeling,
we adopt surface brightness profiles ranging from $n=0.6$ to $n=1$ 
  and an effective radius distribution that matches 
  the effective radius--magnitude relation of the NGC 3585 dwarf galaxy candidates
  in Figure \ref{fig-censurface} (see Section \ref{cmdstructure}). 
Based on reproducing our the visual inspection procedure, 
  we determine 
    that we are 90\% complete down to 
    \textcolor{black}{$I\approx19.8$ mag in the deep field
    ($M_V\approx-11.0$ mag at the distance of NGC 3585), and to $I\approx19.6$ mag in 
    the shallow field.}

\textcolor{black}{
For a galaxy survey that targets galaxies larger than a given radius ($r_{lim}$) with a given surface brightness ($\mu_{lim}$), 
 we can predict detection limits for radius or surface brightness versus total magnitude \citep{fer88,mul15}.
For our dwarf galaxy survey, in which we detect galaxies larger than 10 arcsec ($r_{lim}=5$ arcsec) down to a $\mu_{lim}=28$ mag arcsec$^{-2}$,
  the completeness boundaries are drawn as dotted curves in Figure \ref{fig-censurface}, assuming that the dwarfs have exponential surface brightness profiles.
  Most of our dwarf galaxy candidates are located on the left side of the boundaries as expected.
The results from this calculation are also consistent with those from the injection of artificial galaxies.
}

\textcolor{black}{
Often the central region of a group is contaminated by diffuse intragroup light 
\citep{wat14,wat15, mih17}.
We now check whether diffuse intragroup light is affecting our detection of dwarf galaxy candidates.
We compare the recovery for the inner (within $\sim 0.15$ Mpc, the area that would suffer the most from intragroup light)
and the outer regions. 
The completeness difference is less than 5\%, which is within our $1\sigma$ uncertainty, 
so we conclude that intragroup light is not affecting 
this study.
}

\textcolor{black}{
Finally, our dwarf galaxy candidates could be confused with Galactic cirrus, which is often observed in deep optical images
\citep{miv16}.  
Because Galactic cirrus is attributed to dust grains,
  we examine the $WISE~ 12\micron$ map of this region
  to help us discriminate between dwarf galaxy candidates and Galactic cirrus. 
Our NGC 3585 fields are located outside clear regions of Galactic cirrus that can be seen in the $WISE~ 12\micron$ map. 
In terms of color and surface brightness, Galactic cirrus has red colors
(1.3 mag $< (g-r)_0 < 2.0$ mag and 1.5 mag $< (B-V)_0< 2.2$ mag) \citep{lud12}, while
our dwarf galaxy candidates have relatively blue colors, 0.1 mag $< (B-V)_0 <$ 1.0 mag (see Section \ref{cmdstructure}).
Lastly, all of our dwarf galaxy candidates have a brighter central surface brightnesses than the surface brightness of Galactic cirrus ($\mu_B > 27$ mag arcsec$^{-2}$) \citep{cor10}. 
We conclude that most, if not all, of our dwarf galaxy candidates are not misclassified Galactic cirrus.
}

\subsection{Surface Photometry and Catalog}\label{photometry}

We measure 
the surface brightness profiles of the 46 dwarf galaxy candidates and the 4 previously known dwarf galaxies in $B, V,$ and $I$ using the IRAF/{\it ELLIPSE} task.
\textcolor{black}{
Because many of our candidates are of low surface brightness, 
we do the photometry in two steps.
First, we measure the brightness of the candidates
   without fixing the parameters (e.g. center, position angle, and ellipticity) using the $ELLIPSE$ task
   on the $I$-band image, which is relatively deeper than other band images.
We then fix the parameters to be the values measured around the resulting effective radius and
  measure the surface brightness profile of each candidate on the $BVI$-band images.
We use $1.2$ arcsec linear steps along the semi-major axis and
measure the mean isophotal flux. 
The uncertainty in each measurement is determined from the combination of three sources as follows:
 (1) the mean error of the isophotal flux as given by $ELLIPSE$,
 (2) the variation of sky background level, 
    which is estimated from the fluxes in several of the outermost radial bins,
 and (3) the variation among fitting parameters, which we estimate by 
     calculating several hundred trials while varying the parameters 
     according to the uncertainties of the fixed parameters 
     returned originally by $ELLISPE$.}
Finally, we transform the instrumental surface brightness using the zero points and color terms measured in Section \ref{data}.

In Figure \ref{fig-radimu0a} we show the surface brightness (upper panels) and color (lower panels) profiles of the NGC 3585 dwarf galaxy candidates.
Most candidates have roughly an exponential ($n\sim1$) surface brightness profile and a constant color profile.
The faintest surface brightnesses reach  
  approximately $28$ mag arcsec$^{-2}$. 
%
The weighted mean colors ($\langle\mu_{(V-I)}\rangle$ and $\langle\mu_{(B-V)}\rangle$)
  are denoted by the blue dot-dashed and green dotted lines, respectively, in the color profile panels.
  We adopt the mean color as the color, $(V-I)$ and $(B-V)$, of each dwarf galaxy candidate.
The colors ranges are $ 0.3<(B-V)<1.1$ and $ 0.6<(V-I)<1.3$. 

We use   
the  S{\'e}rsic function
  ($\mu_0+1.0857(r/r_0)^{1/n}$),
  where $\mu_0$, $r_0$, and $n$ are the
  central surface brightness, scale length, and S{\'e}rsic curvature index, respectively, to fit the surface brightness profiles.
The best fitting the S{\'e}rsic functions 
are presented as solid curves in Figure \ref{fig-radimu0a}.
\textcolor{black}{We adopt
  the 1-sigma statistical errors estimated from the non-linear least squares fitting procedure \citep{mar09}
  for the uncertainties of the S{\'e}rsic parameters.
Typical uncertainties for our candidates
 ($\langle \sigma(\mu_{0,I}) \rangle \approx 0.12$ mag arcsec$^{-2}$, $\langle \sigma(r_{0,I}) \rangle \approx 0.59\arcsec$, and
  $\langle \sigma(n_{I}) \rangle \approx 0.13$)
  are plotted on Figure \ref{fig-censurface}.}
%
All of the dwarf galaxy candidates have central surface brightness values brighter than
    $\mu_{0,I}\sim 25.4$, $\mu_{0,V}\sim 26.2$, and $\mu_{0,B}\sim 27.1$ mag arcsec$^{-2}$.
They have S{\'e}rsic scale lengths of $2\arcsec <r_{0,I}<24\arcsec$, and 
   curvature indices of $0.4<n_{I}<2.0$ with a median value of $\sim$0.8.
   The best fits for dwarf galaxy candidates in the overlap regions of N3585-1 and N3585-2 fields (Figure \ref{fig-radec})
 are comparable, 
 showing consistency in our measurements.
We estimate their total magnitudes and effective radii from the extrapolation of the S{\'e}rsic fits
  (see Section 3.1 of \citet{chi09} for the details of the method). 
The faintest total magnitudes in the $B,V,$ and $I$-bands 
  are 22.4, 21.6, and 20.5 mag, respectively.
The smallest effective radius ($r_{e,I}$) in the $I$-band is $\sim 1.5$ arcsec 
  (corresponding to $\sim 150$ pc at the distance of NGC 3585).
\textcolor{black}{
The uncertainties of their total magnitudes and effective radii are derived by measuring several hundred trials while varying according to the uncertainties of the S{\'e}rsic parameters.
The typical values, $\langle \sigma(I) \rangle \approx 0.14$ mag and $\langle \sigma(r_{e,I}) \rangle \approx 0.76\arcsec$, for our dwarf galaxy candidates
   are also plotted on Figure \ref{fig-censurface}.}


In Table \ref{tab-cat} we present the photometric results for 
  the 46 dwarf galaxy candidates in the study 
  and the 4 confirmed dwarfs.  
The first column contains the name.
The second and third columns list the central coordinates. 
The 4th column presents the total $I$-band magnitude based on the S{\'e}rsic fit.
The 5th and 6th columns are the observed $(V-I)$ and $(B-V)$ colors, respectively.
The 7th,  8th,  9th, and 10th columns list the central surface brightness ($\mu_{0,I}$), scale length ($r_{0,I}$), curvature index ($n_{I}$), and $I$-band effective radius  ($r_{e,I}$) derived from the S{\'e}rsic fit, respectively.
%
The `N' and `U' flag in the last column
indicate
whether the galaxy candidate is nucleated (`N') and/or if it can be qualified
as an ultra-diffuse galaxy (`U'). See Section \ref{nucleus} and Section \ref{udg} for the details 
of these classifications. 
%

For the purpose of 
constructing the luminosity function (Section 4.4), we also photometer 
the four bright group galaxies (see Section 3.1 for details).
The resulting photometry of the bright galaxies is as follows:
 NGC 3585 ($M_V = -22.52$), UGCA230 ($M_V = -19.09$),
 ESO 503-G007 ($M_V = -18.70$) and ESO 438-G012 ($M_V = -18.67$).
 These values are within 0.5 mag of the total magnitudes in {\it NED}.

\subsection{Radial Number Density}\label{numden}

In Figure \ref{fig-rnden} we present the 
   dwarf galaxy candidate number density
 as a function of the projected radius ($R$) 
 from the center of NGC 3585. 
The profile declines from the center followed by flattening in the outermost region ($1.5\arcdeg \lesssim R\lesssim 2\arcdeg$).
%
If the flattening represents the distribution of unrelated
background galaxies, then
 the observed number density profile is compatible with
a surface density distribution that is either exponential (Figure \ref{fig-rnden}a) or
power-law (Figure \ref{fig-rnden}b).
For the exponential form, the best-fit function is $\Sigma = e^{-1.95(\pm0.67) R + 3.31(\pm0.55)} + 1.80(\pm0.50)$ deg$^{-2}$,
while for the power-law form, it is $\Sigma = 2.90(\pm0.93)R^{-1.52(\pm0.29)}+1.80$ deg$^{-2}$.
Based on the former, 
  we estimate the background contamination to be 
 ten objects (23\% of our total) within a 1.5 deg radius of NGC 3585.
We believe that the remaining four dwarf galaxy candidates located beyond $R = $ 1.5 
are most likely 
background sources.  
\textcolor{black}{
The contamination level we derive from the exponential model, 1.80 deg$^{-2}$, 
  is consistent with the contamination value, 1.75 deg$^{-2}$, estimated from the background field counts
  (see Section \ref{bgtest} in detail).
}

\textcolor{black}{
We considered the possibility of using surface brightness fluctuation (SBF) measurement to ascertain membership, 
but considering our observing conditions (1.3 arcsec seeing in $I$-band), the target's distance (20 Mpc), and the exposure time (10 hours).
Comparing to some work in this area (e.g. \citet{mie03, car19b}), we confirmed out expectation that this is not
possible.
}

\subsection{Color and Structure Parameters}\label{cmdstructure}

We present the color-magnitude diagrams for the dwarf galaxy candidates
  in the NGC 3585 field in Figure \ref{fig-cmd}.
The mean $(B-V)_0$ and $(V-I)_0$ of the dwarf galaxy candidates corrected for Galactic extinction  are
  $0.69\pm0.18$ and $0.87\pm0.14$, respectively. 
These mean values are similar to those obtained for other groups:
  the M106 group, $\langle (B-V)_0\rangle\approx 0.73$ \citep{kim11},
  the NGC 2784 group, $\langle (B-V)_0\rangle\approx 0.67$, $\langle (V-I)_0\rangle\approx 0.85$ \citep{par17}, 
  and the M83 group,  $\langle (B-V)_0\rangle\approx 0.82$ \citep{mul15}. 
When 
  the colors of the dwarf galaxy candidates are overlaid with those of dwarf galaxies in other groups (Figure \ref{fig-cmd}), we find that 
the color distributions of the NGC 3585 dwarf galaxy candidates are consistent with those of other groups 
(e.g. the M83 group from \citealt{mul15}, the NGC 2784 group from \citealt{par17}). 
In general, the early-type galaxies in galaxy clusters have a color-magnitude relation:
 the brighter the galaxies, the redder they appear (e.g. the Virgo cluster from \citealt{lis08}
  and the Ursa Major cluster from \citealt{pak14}). 
  The dwarf galaxy candidates in the NGC 3585 also follow the relation as shown in Figure \ref{fig-cmd}.
Using dwarf galaxy candidates brighter than $M_V=-11$ mag in the NGC 3585 group from this study and
 the NGC 2784 group from \citet{par17}, 
 the dwarf galaxy candidates follow color--magnitude relations:
 $(B-V)_0 = -0.012 ~ M_V + 0.518$ and 
 $(V-I)_0 = -0.013 ~ M_V + 0.691$. 
These results for the color distribution and relation indicate that many of the dwarf galaxy candidates in the NGC 3585 field may be early-type galaxies.

In Figure~\ref{fig-censurface} we compare the structural parameters, $V$-band
central surface brightnesses ($\mu_{0}$), effective radii ($r_{e}$), and S{\'e}rsic $n$ indices, of the dwarf galaxy candidates in the NGC 3585 field
to those found in other groups.
The $\mu_{0,V}$ and $M_V$ values are derived from the estimated $\mu_{0,I}$ and $M_I$ (Section 3.2)
using the $(V-I)$ color of the dwarf galaxy candidates assuming that they are at the
same distance of NGC~3585. The $r_{e}$ and $n$ values are those obtained from fits to the $I$-band images.  
The values for the structural parameters of dwarf galaxies in other groups 
are compiled from the literature:
  \citet{mul15} for the M83 group,
  \citet{chi09} for the M81 group, and
  \citet{par17} for the NGC 2784 group.
As seen in the M83 group \citep{mul15} and the NGC 2784 group \citep{par17}, 
the central surface brightnesses and the effective radii of the dwarf galaxy candidates in the NGC 3585 group
increase with galaxy luminosity.
In the case of the S{\'e}rsic curvature index,
the median value, $n\approx 0.8\pm 0.1$, of the dwarf galaxy candidates in the NGC 3585 field is similar to that of dwarf galaxies in other groups:
  $n\approx 1.1\pm 0.3$ for the M83 group \citep{mul15},
  $n\approx 0.6\pm 0.2$ for the M81 group \citep{chi09}, and
  $n\approx 0.8\pm 0.1$ for the NGC 2784 group \citep{par17}. 
The properties of the NGC 3585 dwarf galaxy candidates are consistent with those of dwarf galaxies in other groups. 
  
\textcolor{black}{
\subsection{Estimation of Contamination Level by Background Diffuse Sources}\label{bgtest}
%
In order to estimate the level of contamination of the dwarf galaxy candidates in NGC 3585 fields identified in Section \ref{dwsearch} by sources in background, 
we adopt the KK196 field as a control field. 
The KK 196 field (R.A.$_{J2000}=13^h21^m47.42^s$, Decl.$_{J2000}=-45\arcdeg 03\arcmin 46.2\arcsec$), which is about 30 degrees from that of NGC 3585, has been observed together with the NGC 3585 field in our KSP program, so that both the fields have almost identical integrated exposure times and depth in stacked images. In addition, the KK 196 field belongs to Centaurus~A group at 4 Mpc distance, and this makes the use of the field as a control field more reliable because the dwarf galaxies in this field associated with the Centaurus~A group can easily be resolved due to their proximity. Our three classifiers searched for dwarf galaxy candidates in the KK196 field in
the same manner we described in Section 3.1. As a result, we found 4 resolved dwarf galaxy candidates that belong Centaurus~A group -- two of them were previously identifed as KK196 and KK203, while the rest two are newly indentified in this study -- as well as 7 unresolved dwarf galaxy candidadtes that appear to be unassociated with Centaurus~A group. We, therefore, consider these 7 candidates as background dwarf galaxies. 
}

We now carry out the same analysis on these seven candidates as we did on the NGC 3585 dwarf galaxy candidates.
Results will be reported in greater detail by Park et al. (2019, in preparation). 
We summarize our results with a focus on understanding the background contamination in the NGC 3585 field as follows:
 \textcolor{black}{
 (1) the spatial distribution of the 7 observed diffuse candidates is compatible with 
random distribution from both the 1 and 2-dimensional Kolmogorov-Smirnov tests \citep{pre88},} 
 (2) assuming that these objects are at the distance of NGC 3585,
   their distribution in the color-magnitude diagram (mean colors $(B-V)_0=0.67\pm0.08$ and $(V-I)_0=0.82\pm0.10$ and  $-13.6<M_V<-11.1$ mag)
   and their S{\'e}rsic structural parameters (median $n \approx 0.7\pm 0.2$,)
    are consistent with those of the NGC 3585 dwarf galaxy candidates,
 (3) their projected number density is about 7/(2 deg $\times$ 2 deg),
   or $1.75(\pm0.66)$ deg$^{-2}$, which is very similar to the value estimated from the number density profile of NGC 3585 dwarf galaxy candidates (1.80 deg$^{-2}$, see the Section \ref{numden}), and 
 (4) none of these candidates have either a nuclear source or are UDGs. While the color and magnitude similarities make it difficult to reject this type of background source, we have a robust estimate of their contribution to the satellite number density profiles and an indication that
 one way to identify bonafide dwarf galaxies may be if they are either nucleated (see Section \ref{nucleus}) or ultra-diffuse (see Section \ref{udg}).

\section{Nature of the NGC 3585 Group}\label{discuss} 

\subsection{Radial Distributions} 

In clusters and groups, the more massive galaxies tend to be centrally concentrated
\citep{pre12, rob15}.
To investigate if dwarf galaxies in the NGC 3585 group conform to this trend,
 we divide our sample in two: bright ($M_V<-12.5$ mag) and faint ($M_V\geq-12.5$ mag).
We show the radial number density profiles for these two subsamples in Figure \ref{fig-radcoldenudg}(a).  %
The best-fit exponential functions
for the bright and faint populations are 
$\ln(\Sigma) \simeq\ -0.58 (\pm 0.41) R + 0.97 (\pm 0.34)$ and 
$\ln(\Sigma) \simeq\ -0.76 (\pm 0.08) R + 1.26 (\pm 0.07)$, 
respectively. 
Given the uncertainties in the fits, 
both populations of dwarf galaxy candidates show
similar density profiles. Thus, mass segregation 
is not apparent among the dwarf galaxy candidates 
in the NGC 3585 group.
In Figure \ref{fig-radcoldenudg}(b) we plot the mean color
of the dwarf galaxies in NGC 3585 vs. projected radius from the group center. 
There is no apparent color variation
with radius.

%
The mass and color radial behavior of the dwarf galaxy candidates in the NGC 3585 differ from the significant radial dependencies that we found in 
the NGC 2784 group \citep{par17}. 
This distinction 
suggests 
that the NGC 3585 group \textcolor{black}{might be} a dynamically younger system where mass segregation has not yet developed.

\subsection{Nucleated Dwarf Galaxies}\label{nucleus}

Some dwarf galaxies have a distinct nucleated source in their central region \citep{cot06,tre09,tur12} and the dwarf galaxy candidates s in the NGC 3585 group are no exception.
In Table \ref{tab-cat} we classify eight dwarf galaxy candidates among the 50 in the NGC 3585 group 
  as nucleated  based on the images
  (Figure \ref{fig-graymapa}) and
  the surface brightness profiles (Figure \ref{fig-radimu0a}).
\textcolor{black}{
These classified nucleated candidates have an evident point source that is distinctly brighter (approximately $\gtrsim0.5$ mag arcsec$^{-2}$) than central surface brightness estimated by S{\'e}rsic fit to the entire galaxy.}
The incidence of the nucleated dwarf galaxy candidates is 20\% (8 out of 50) in the NGC 3585 group, which 
 is consistent with that found in other groups (20\% in the Local Group early-type \citep{tur12} and 
10-30\% for several nearby galaxy groups \citep{tre09}, and 10\% in 
the NGC 2784 group \citep{par17}\footnote{
  There are four nucleated dwarf galaxy candidates (KSP-DW13, NGC 2784 dw01, KK72, and KSP-DW15) in the NGC 2784 group.
  }
  ). 
In contrast, 
%
in the Virgo and Fornax clusters studies have found 
that about 60--80\% of the dwarf galaxies have nuclei \citep{cot06,lis07,tur12, geo14}. 
Although selection effects could play a role, 
cluster samples are biased to brighter samples, 
  even our bright group sample has a small nucleated incidence ($\sim 30$\%; Figure \ref{fig-nuc}a).
Unfortunately, the significance of the group results 
  is undermined by small number statistics. Confirmation of the difference between groups and clusters awaits larger samples.
\textcolor{black}{
  Nevertheless, none of the seven dwarf galaxy candidates in the control field (see Section \ref{bgtest})   
  has a nucleus, which indirectly supports a conjecture that the nucleated dwarf galaxy candidates 
  are associated with the NGC 3585 group.
}

In Figure \ref{fig-nuc}, we compare the luminosities ($M_V$), colors ($(B-V)_0$), radial distribution from the group center ($R$), 
  central surface brightness ($\mu_{0,V}$), effective radii ($r_e$), and S{\'e}rsic-$n$'s of the nucleated dwarf galaxy candidates to those of other dwarf galaxy candidates.
We overlay 
 the distribution of the number ratio ($N_{NUC}/N_{ALL}$) 
 in each panel.
From these plots, we conclude that  
   nucleated dwarf galaxy candidates are more common \textcolor{black}{among brighter (a, d), redder (b), larger (e), and more centrally concentrated (f) dwarf galaxy candidates.} 
We note that due to the relationships among these parameters, a high incidence for redder dwarf galaxy candidates is related 
in corresponding higher incidence among brighter, more concentrated dwarf galaxy candidates 
  due to the color--magnitude relation and 
  the magnitude--structural parameter trends 
  described in Section 3.4.
The correlation between nuclear sources and color is consistent with those obtained for dwarf elliptical galaxies (dEs) in nearby galaxy clusters
\citep{rak04, lis07}.
However, the incidence of nucleated dwarf galaxy candidates does not vary with distance from the center in the NGC 3585 group
(Figure \ref{fig-nuc}c)
which is in contrast to the trend in clusters. The nucleated dwarfs in the Virgo cluster are located in denser environments.
 If the radial trend in Virgo is due to mass segregation, with the nucleated dwarfs being the more massive and sinking toward the center, our result
 might again suggest that the NGC 3585 system \textcolor{black}{might be} dynamically young.
%

\subsection{Ultra-Diffuse Galaxies}\label{udg}

Numerous ultra-diffuse galaxies (UDGs) have recently been found in nearby galaxy clusters and groups 
\citep{van15, vdb16,mul18, zar19}.
We find four UDG candidates in the NGC 3585 group based on
  central surface brightness ($\mu_{0,V}\gtrsim 23.7$ mag arcsec$^{-2}$) and 
  effective radius ($r_{e,V}\gtrsim 1.5$ kpc) criteria 
  \citep{van15}.
The UDG candidates in the NGC 3585 group (see Table \ref{tab-cat})\footnote{ 
  The NGC 2784 group \citep{par17}
  has just one UDG candidate (NGC2784dw01)
  located near the group center ($R=0.01$ Mpc)
  with $\mu_{0,V}=24.4$ mag arcsec$^{-2}$ and $(B-V)_{0}=0.8$,
  but the candidate can only marginally be classified as a UDG
  due to its low effective size of $r_{e,V}=1.3$ kpc. }
  have parameters in the range of $24.3<\mu_{0,V}<25.9$ mag arcsec$^{-2}$, 
  $1.6< r_{e,V}< 2.0$ kpc.
 $0.55\le (B-V)_{0}\le 0.75$, and $n_{V}\lesssim 1.0$.
 The latter two quantities are
  similar to the average values, $(B-V)_{0}\approx 0.69$ and $n\approx 0.8$, of all the dwarf galaxy candidates in the NGC 3585 group.
\textcolor{black}{The absence of UDGs among the control field dwarf candidates (see Section \ref{bgtest}) indirectly supports the conjecture
   that the UDG candidates are associated with the NGC 3585 group.
}

An interesting properties of the four UDGs in this group is that they are
centrally concentrated, all
within  $R<0.15$ Mpc
  (Figure \ref{fig-radcoldenudg}(b)).
%
A strong 
central concentration is in conflict with results from UDGs in galaxy clusters
  \citep{van15,vdb16,rom17a} and some galaxy groups \citep{mer16, rom17b, mul18}, which
  instead find a deficit of UDGs in the central region.
 While the statistical significance of the result for NGC 3585 is again limited due to small numbers, the result again suggests that 
the NGC 3585 group might be a dynamically younger system in which centrally located UDG candidates have not yet been tidally disrupted.
Finally, we add our measurement to previous ones to investigate
 the number abundance of UDGs as a function of group mass.
  In this instance, we consider a velocity dispersion as a system mass. 
A value of 4 UDGs in a system with a velocity dispersion of $\sigma_{\upsilon}=70$ km s$^{-1}$ (see Section 3.1),
 causes a flattening of the relationship between the UDG number abundance and system mass (see 
 Figure 6 by \citet{rom17b} or Figure 10 by \citet{lee17}). If confirmed for similar low mass groups, 
 this result may indicate that UDGs form preferentially in groups \citep{rom17b} or are less effectively destroyed.
 
\subsection{Comparison with Luminosity Functions of Other Groups}

To compare the luminosity function (LF) of the galaxies in the NGC 3585 group with
  those of other galaxy groups, 
we use cumulative LFs 
because, in general, the number of galaxies in a group is small \citep{tre09}.
To construct the LF for the NGC 3585 group,
 we only use galaxies with $R<0.5$ Mpc 
   to minimize the effect of background contamination (Section 3.3). 
The other groups in this comparison include 
the M81 group \citep{chi09}, 
the M83 group \citep{mul15}, 
\textcolor{black}{ 
the M96 group \citep{mul18},
the M101 group \citep{dan17, ben17,ben19, car19a}},
the M106 group \citep{kim11},
the NGC 2784 group \citep{par17}, and
the NGC 5128 group \citep{tul15,crn16}. 
The $R < 0.5$ Mpc condition is satisfied for all of these other groups except the NGC 5128 group. However, in that case most of the dwarfs are spectroscopically confirmed members so contamination is not an issue.
In Figure \ref{fig-clfalpha}(a) we compare the cumulative LFs of the NGC 3585 group to those of these \textcolor{black}{seven} other groups.

 The \textcolor{black}{eight} groups have roughly similar LF shapes and faint-end slopes.
 However, 
 we do notice that certain groups (M81, M83, NGC 2784, and NGC 3585) have a clear faint-end hump,  a steep increasing of the cumulative numbers around $M_{V}\approx -13$ mag.
 These LF humps could be a real structure, with physical meaning that could be exploited, or
 perhaps the result of background contamination. To resolve this ambiguity requires either deep spectroscopic surveys to determine membership or a larger, homogeneous photometric survey, such as our complete KMTNet dwarf galaxy search project. With a much larger sample it would be possible to establish whether this is a universal feature and whether it is always observed at the same apparent magnitude, in which case one would suspect it is the result of background contamination, or at the same absolute magnitude, in which case one would suspect a physical effect.
\textcolor{black}{
 We find from the background contamination test (Section \ref{bgtest}) that contaminating systems tend to be 
    fainter than $M_{V}\approx -13$ mag and so 
     the hump in the case of the NGC 3585 group could be at least partially attributed to contamination.}

We fit the observed cumulative LF of each group, for $M_V<-10$ mag,
 to a cumulative Schechter function \citep{sch76}
 using the technique described in \citet{chi09}. 
The best-fit faint-end slope, $\alpha=-1.39 \pm 0.03$,
  for the cumulative LF of the NGC 3585 group 
   is similar to those of other groups (Figure \ref{fig-clfalpha}a).
The results for the cumulative LFs of the other groups \textcolor{black}{except the M96 group with $\alpha=-1.36 \pm 0.03$}
  are in detail introduced in Section 4.2 by \citet{par17}
  \textcolor{black}{and the M101 group is excluded for further comparison 
   because the group shows an abnormal LF shape: very flat for $M_V<-10$ mag and very steep for $M_V>-10$ mag 
  (see \citet{ben19} in more detail).}
We find that the groups have a range of slopes, with the LF slope 
  of the LG
 \citep[$\alpha\approx -1.0$;][]{chi09, mcc09, kim11, fer16, par17}, being considerably flatter than any of these. On the other hand, all observed faint-end slopes are
    much flatter than that of the subhalo mass function in the $\Lambda$CDM model ($\alpha=-1.8$) \citep{tre02}.
  %

To search for a dependence of the faint-end slope on group properties,
  we initially investigated a correlation between group mass and LF slope
  in our previous study \citep{par17}.
  There we \textcolor{black}{suggested} that the LF slope flattens as group mass increases.
  However, 
  the small sample size, the large uncertainties, \textcolor{black}{and the membership problem}
    render this only as a preliminary result.
%
Including the new results for the NGC 3585 group \textcolor{black}{and the M96 group} 
in Figure \ref{fig-clfalpha} seems to support the previous claim, although we still need more data.
To construct this figure, we adopted 
    $\sigma_{group}$ as an indicator for group mass. 
    The values of $\sigma_{group}$ for groups other than NGC 3585 \textcolor{black}{and M96} are from Table 4 in \citet{par17}.
 The values for the NGC 3585 group, 
  $\sigma_{group}=70$ km s$^{-1}$, 
  \textcolor{black}{ and the M96 group, $\sigma_{group}=233$ km s$^{-1}$ listed as M105, are} from \citet{mak11}. 
%
The relation between $\alpha$ and $\sigma_{group}$ can be expressed as 
  \textcolor{black}{$\alpha=0.0003 ~\sigma_{group}-1.353$}, 
  but a correlation analysis suggests that the confidence with which we can claim a correlation is only at \textcolor{black}{about $1\sigma$}. 
This marginal nature of the result 
could 
be the result of systematic differences among group results obtained from diverse surveys 
  and of small sample statistics
  \textcolor{black}{as well as of the group membership}.
A large and uniform survey is required to establish this result.
The homogeneous samples obtained from the KMTNet survey of dwarf galaxies \textcolor{black}{in $\gtrsim 30$ groups}
  will serve this purpose.

We close this section by noting the apparent discrepancy between our results and those of \cite{zab00} and \cite{bal01}, who find steeper faint end slopes in denser, more massive environments. 
  However, those previous results characterize significantly more luminous galaxies than those characterized here (e.g. $-17.8 < M_R < -19.8$, for \cite{zab00}). If the faint-end LF is not a simple power-law, as suggested earlier in this section, then constraining $\alpha$ over different absolute magnitude ranges will lead to apparently conflicting results. Interpreting differences among results from various studies is also complicated by the comparison of LFs derived from spectroscopically confirmed samples, such as that used by \cite{zab00}, and those derived from samples that either apply statistical background corrections or argue that they are minor. It is therefore premature to place much weight on the trend seen in Figure \ref{fig-clfalpha} until the larger group survey securely establishes the shape of the faint-end LF and the role of background contamination.

\section{Conclusion and Summary}\label{summary}

We identify a total of 46 new dwarf galaxy candidates in the NGC 3585 group over 7 $\square^\circ$ 
using  wide and deep images obtained from the  KMTNet Supernova Program. 
We present $BVI$ surface photometry 
  for 50 dwarf galaxy candidates, including 4 previously known dwarf galaxies. 
For various reasons, including the radial distributions of bright and faint, nucleated and non-nucleated, standard and ultra-diffuse galaxies, we suggest that the NGC 3585 group may be a dynamically younger system than the typical group.
We summarize our results as follows.

\begin{enumerate}

\item
There is a significant population of dwarf galaxies in the NGC 3585 group.
The projected number density of dwarf galaxy candidates decreases exponentially 
(or with power index, $\alpha \approx -1.5$) with distance from the center of NGC 3585 and
flattens beyond 1.5 degree ($\sim$ 0.5 Mpc).
The background contamination estimated from this density profile is about 23\%,
which implies that about 36 of the 46 dwarf galaxy candidates within 1.5 degree are members of the NGC 3585 group.
\textcolor{black}{This background level is similar to the value based on the control field.}

\item 
There is nothing unusual about the internal structure of the NGC 3585 dwarf galaxy candidates.
They have  
  color and S{\'e}rsic $n$ parameter distributions 
 that are consistent with those of dwarfs in other galaxy groups (M81, M83, and NGC 2784). They also have a color-magnitude relation similar to the early-type galaxies in galaxy clusters.

\item  
There is an unusual lack of radial dependence in the properties of the NGC 3585 dwarf galaxy candidates.
The  color and the number density profiles do not show a radial dependency. 
We interpret
this result to mean that the NGC 3585 group \textcolor{black}{might be} dynamically younger than at least the NGC 2784 group in our previous study.

\item
The incidence of nucleated dwarfs may offer an insight into the role of environment in galaxy evolution. We find
eight nucleated dwarf galaxy candidates in the NGC 3585 group. The incidence of nucleation appears to be   
  larger at brighter magnitudes and redder colors. We find no radial dependence on the incidence. 
The incidence of the nucleated dwarf galaxy candidates in this group is roughly 20\%, which is much lower than that in galaxy clusters (60 -- 80\%), and roughly consistent with what is
found in other groups \citep{par17}

\item
The radial distribution of ultra-diffuse galaxies may also be a promising clue.
The four UDG candidates we identify in the NGC 3585 group are all within the central region of the group, in contrast with what is found in clusters 
\citep{van15,vdb16}. Perhaps, if this group is dynamically young, the UDGs have not yet suffered  sufficiently from tidal disruption. 

\item 
The faint-end slope, $\alpha$, of the galaxy luminosity function (LF) correlates weakly with group mass. For NGC 3585 we measure 
$\alpha \approx -1.39\pm0.03$ from the cumulative LF. 
The faint-end slopes of the LFs \textcolor{black}{seem to} become flatter as the group masses increase.
However, \textcolor{black}{even when group membership problem is excluded,}
this correlation is still not highly statistically significant and systematic uncertainties in the shape of the LF remain. 
A large, homogeneous survey for nearby groups, 
such as our KMTNet dwarf search project, should establish or refute the existence of this correlation.

\end{enumerate}


\acknowledgments 

The authors are grateful to the staffs of the KMTNet. 
This research has made use of the KMTNet system
operated by the Korea Astronomy and Space Science Institute
(KASI), and the data were obtained at three host sites of CTIO
in Chile, SAAO in South Africa, and SSO in Australia.
H.S.P. was supported in part by the National Research Foundation of Korea (NRF) grant
funded by the Korea government (MSIT, Ministry of Science and ICT) (No. NRF-2019R1F1A1058228).
D.-S.M. was supported in part by a Leading Edge Fund from the
Canadian Foundation for Innovation (project No. 30951) and a
Discovery Grant from the Natural Sciences and Engineering
Research Council of Canada.



\clearpage

\begin{deluxetable}{ccccccc}
\tabletypesize{\tiny}
\tablewidth{0pc}
\tablecaption{Observing Log for Stack Images\label{tab-obslog}}
\tablehead{
\colhead{Field} &
\colhead{Filter} &
\colhead{R.A.(J2000)} &
\colhead{Decl.(J2000)} &
\colhead{T(exp)} &
\colhead{Date(UT)} &
\colhead{Seeing} \\
\colhead{} &
\colhead{} &
\colhead{(hh:mm:ss)} &
\colhead{(dd:mm)} &
\colhead{(N$\times$sec)} &
\colhead{(yyyy.mm.dd.)} &
\colhead{(arcsec)} 
}
\startdata
N3585-1-Q0 & $I$ & 11:13:37 & -25:41 & 596 $\times$  60 s & 2016.01.21. -- 2017.02.16. &  1.3 \\ 
N3585-1-Q0 & $V$ & 11:13:37 & -25:41 & 558 $\times$  60 s & 2016.01.21. -- 2017.02.15. &  1.4 \\ 
N3585-1-Q0 & $B$ & 11:13:37 & -25:41 & 525 $\times$  60 s & 2016.01.21. -- 2017.02.15. &  1.5 \\ 
N3585-1-Q1 & $I$ & 11:08:53 & -25:41 & 606 $\times$  60 s & 2016.01.21. -- 2017.02.15. &  1.3 \\ 
N3585-1-Q1 & $V$ & 11:08:53 & -25:41 & 568 $\times$  60 s & 2016.01.21. -- 2017.02.15. &  1.4 \\ 
N3585-1-Q1 & $B$ & 11:08:53 & -25:41 & 525 $\times$  60 s & 2016.01.21. -- 2017.02.15. &  1.5 \\ 
N3585-1-Q2 & $I$ & 11:13:39 & -26:48 & 601 $\times$  60 s & 2016.01.21. -- 2017.02.16. &  1.3 \\ 
N3585-1-Q2 & $V$ & 11:13:39 & -26:48 & 566 $\times$  60 s & 2016.01.21. -- 2017.02.15. &  1.4 \\ 
N3585-1-Q2 & $B$ & 11:13:39 & -26:48 & 538 $\times$  60 s & 2016.01.21. -- 2017.02.15. &  1.5 \\ 
N3585-1-Q3 & $I$ & 11:08:51 & -26:48 & 590 $\times$  60 s & 2016.01.21. -- 2017.02.16. &  1.3 \\ 
N3585-1-Q3 & $V$ & 11:08:51 & -26:48 & 545 $\times$  60 s & 2016.01.21. -- 2017.02.15. &  1.4 \\ 
N3585-1-Q3 & $B$ & 11:08:51 & -26:48 & 516 $\times$  60 s & 2016.01.21. -- 2017.02.15. &  1.4 \\ 
N3585-2-Q0 & $I$ & 11:17:38 & -26:31 & 274 $\times$  60 s & 2017.02.16. -- 2017.06.24. &  1.3 \\ 
N3585-2-Q0 & $V$ & 11:17:38 & -26:31 & 259 $\times$  60 s & 2017.02.16. -- 2017.06.24. &  1.4 \\ 
N3585-2-Q0 & $B$ & 11:17:38 & -26:31 & 246 $\times$  60 s & 2017.02.16. -- 2017.06.24. &  1.5 \\ 
N3585-2-Q1 & $I$ & 11:12:52 & -26:31 & 271 $\times$  60 s & 2017.02.16. -- 2017.06.24. &  1.3 \\ 
N3585-2-Q1 & $V$ & 11:12:52 & -26:31 & 249 $\times$  60 s & 2017.02.16. -- 2017.06.24. &  1.3 \\ 
N3585-2-Q1 & $B$ & 11:12:52 & -26:31 & 239 $\times$  60 s & 2017.02.16. -- 2017.06.24. &  1.4 \\ 
N3585-2-Q2 & $I$ & 11:17:40 & -27:38 & 273 $\times$  60 s & 2017.02.16. -- 2017.06.24. &  1.2 \\ 
N3585-2-Q2 & $V$ & 11:17:40 & -27:38 & 258 $\times$  60 s & 2017.02.16. -- 2017.06.24. &  1.3 \\ 
N3585-2-Q2 & $B$ & 11:17:40 & -27:38 & 246 $\times$  60 s & 2017.02.16. -- 2017.06.24. &  1.4 \\ 
N3585-2-Q3 & $I$ & 11:12:50 & -27:38 & 273 $\times$  60 s & 2017.02.16. -- 2017.06.24. &  1.2 \\ 
N3585-2-Q3 & $V$ & 11:12:50 & -27:38 & 243 $\times$  60 s & 2017.02.16. -- 2017.06.24. &  1.3 \\ 
N3585-2-Q3 & $B$ & 11:12:50 & -27:38 & 228 $\times$  60 s & 2017.02.16. -- 2017.06.24. &  1.4 \\ 
\enddata
\end{deluxetable}

\begin{deluxetable}{cccccc}
\tabletypesize{\scriptsize}
\tablewidth{0pc}
\tablecaption{\textcolor{black}{Standardization of Instrumental Magnitudes}\label{tab-stdmag}}
\tablehead{
\colhead{Field} &
\colhead{$c(B-V)$} &
\colhead{$zero(B)$} &
\colhead{$rms(B)$} &
\colhead{$zero(V)$} &
\colhead{$zero(I)$} \\
\colhead{} &
\colhead{} &
\colhead{(mag)} &
\colhead{(mag)} &
\colhead{(mag)} &
\colhead{(mag)} 
}
\startdata
N3585-1-Q0 &   0.32 $\pm$ 0.02 &  28.09 $\pm$ 0.01 &   0.04 &  28.11 $\pm$ 0.03 &  28.01 $\pm$ 0.05 \\
N3585-1-Q1 &   0.28 $\pm$ 0.01 &  28.15 $\pm$ 0.01 &   0.02 &  28.12 $\pm$ 0.03 &  28.06 $\pm$ 0.04 \\
N3585-1-Q2 &   0.28 $\pm$ 0.01 &  28.17 $\pm$ 0.01 &   0.02 &  28.14 $\pm$ 0.03 &  28.04 $\pm$ 0.08 \\
N3585-1-Q3 &   0.26 $\pm$ 0.02 &  28.14 $\pm$ 0.01 &   0.03 &  28.11 $\pm$ 0.04 &  28.05 $\pm$ 0.06 \\
N3585-2-Q0 &   0.25 $\pm$ 0.02 &  28.26 $\pm$ 0.02 &   0.03 &  28.19 $\pm$ 0.04 &  28.12 $\pm$ 0.07 \\
N3585-2-Q1 &   0.27 $\pm$ 0.01 &  28.26 $\pm$ 0.01 &   0.02 &  28.20 $\pm$ 0.03 &  28.06 $\pm$ 0.07 \\
N3585-2-Q2 &   0.27 $\pm$ 0.01 &  28.22 $\pm$ 0.01 &   0.02 &  28.18 $\pm$ 0.03 &  28.07 $\pm$ 0.04 \\
N3585-2-Q3 &   0.27 $\pm$ 0.01 &  28.20 $\pm$ 0.01 &   0.03 &  28.17 $\pm$ 0.03 &  28.16 $\pm$ 0.08 \\
\enddata
\end{deluxetable}
\clearpage

\begin{deluxetable}{lccccccrrrc}
\tabletypesize{\tiny} 
\tablewidth{0pc}
\tablecaption{Photometric Catalog for Dwarf Galaxy Candidates in the NGC 3585 field\label{tab-cat}}   
\tablehead{
\colhead{ID$~^a$} &
\colhead{R.A.(J2000)} &
\colhead{Decl.(J2000)} &
\colhead{$I~^b$} &
\colhead{$(V-I)~^c$} &
\colhead{$(B-V)~^c$} &
\colhead{$\mu_{0,I}~^d$} &
\colhead{$r_{0,I}~^d$} &
\colhead{$n_{I}~^d$} &
\colhead{$r_{e,I}~^d$} & 
\colhead{flag$~^e$} 
\\
\colhead{} &
\colhead{(hh:mm:ss)} &
\colhead{(dd:mm:ss)} &
\colhead{(mag)} &
\colhead{(mag)} &
\colhead{(mag)} &
\colhead{(mag arcsec$^{-2}$)} &
\colhead{(arcsec)} &
\colhead{} &
\colhead{(arcsec)} & \colhead{}
}
\startdata
    ESO502-G018 & 11:07:18.1 & -25:34:23 &  14.01 &   0.98 &   0.83 &  22.35 $\pm$   0.04 &  23.57 $\pm$   0.68 &   0.70 $\pm$   0.03 &   25.72 &   \\
       KSP-DW32 & 11:07:29.4 & -25:53:50 &  18.29 &   1.00 &   0.83 &  23.74 $\pm$   0.08 &   5.85 $\pm$   0.50 &   0.78 $\pm$   0.09 &    7.71 &   \\
       KSP-DW33 & 11:08:40.8 & -25:52:02 &  19.48 &   0.93 &   0.81 &  23.56 $\pm$   0.15 &   3.40 $\pm$   0.40 &   0.66 $\pm$   0.12 &    3.39 &   \\
       KSP-DW34 & 11:08:42.3 & -25:51:26 &  18.01 &   0.75 &   0.63 &  22.98 $\pm$   0.06 &   5.36 $\pm$   0.27 &   0.57 $\pm$   0.05 &    3.78 &   \\
       KSP-DW35 & 11:09:02.3 & -25:57:16 &  19.02 &   0.85 &   0.83 &  23.53 $\pm$   0.11 &   4.12 $\pm$   0.38 &   0.66 $\pm$   0.12 &    4.12 &   \\
       KSP-DW36 & 11:10:14.2 & -26:27:24 &  20.13 &   0.90 &   0.47 &  23.65 $\pm$   0.08 &   2.29 $\pm$   0.31 &   0.85 $\pm$   0.20 &    3.32 &   \\
       KSP-DW37 & 11:10:36.1 & -26:06:59 &  18.99 &   0.88 &   0.92 &  23.34 $\pm$   0.29 &   2.88 $\pm$   0.84 &   1.03 $\pm$   0.25 &    4.90 & N \\
       KSP-DW38 & 11:10:42.7 & -27:01:15 &  20.17 &   0.80 &   0.82 &  23.69 $\pm$   0.05 &   2.74 $\pm$   0.15 &   0.58 $\pm$   0.09 &    2.02 &   \\
       KSP-DW39 & 11:10:46.2 & -26:31:31 &  18.78 &   1.12 &   0.73 &  24.55 $\pm$   0.15 &   7.65 $\pm$   0.91 &   0.60 $\pm$   0.22 &    6.21 &   \\
       KSP-DW40 & 11:10:48.5 & -25:15:32 &  16.38 &   0.92 &   0.63 &  21.94 $\pm$   0.04 &   5.48 $\pm$   0.29 &   0.93 $\pm$   0.04 &    8.68 &   \\
GALEXASCJ & 11:10:51.5 & -25:44:58 &  14.19 &   0.98 &   0.81 &  19.15 $\pm$   0.15 &   2.29 $\pm$   0.31 &   1.49 $\pm$   0.06 &    4.40 &   \\
    ESO438-G010 & 11:10:51.8 & -27:53:51 &  13.08 &   0.88 &   0.71 &  20.05 $\pm$   0.02 &   8.72 $\pm$   0.19 &   1.13 $\pm$   0.01 &   15.62 & N \\
       KSP-DW41 & 11:11:24.1 & -27:58:09 &  18.35 &   0.85 &   0.73 &  23.52 $\pm$   0.12 &   5.34 $\pm$   0.54 &   0.73 $\pm$   0.11 &    6.39 & N \\
       KSP-DW42 & 11:11:26.1 & -26:32:54 &  18.14 &   0.96 &   0.87 &  24.89 $\pm$   0.08 &  12.65 $\pm$   0.63 &   0.49 $\pm$   0.07 &    5.24 &   \\
       KSP-DW43 & 11:11:44.4 & -26:48:38 &  18.60 &   0.84 &   0.95 &  24.83 $\pm$   0.11 &   9.40 $\pm$   0.77 &   0.60 $\pm$   0.15 &    7.63 &   \\
       KSP-DW44 & 11:11:57.2 & -26:54:59 &  15.16 &   1.05 &   0.95 &  21.99 $\pm$   0.03 &   9.92 $\pm$   0.33 &   0.92 $\pm$   0.03 &   15.58 & N \\
       KSP-DW45 & 11:12:02.7 & -26:25:44 &  20.45 &   0.92 &   1.00 &  24.07 $\pm$   0.78 &   2.42 $\pm$   1.55 &   0.84 $\pm$   0.59 &    3.44 &   \\
       KSP-DW46 & 11:12:13.3 & -27:14:17 &  18.03 &   0.83 &   0.51 &  23.14 $\pm$   0.16 &   5.09 $\pm$   0.75 &   0.76 $\pm$   0.14 &    6.45 &   \\
       KSP-DW47 & 11:12:16.3 & -26:11:16 &  16.23 &   1.05 &   0.80 &  23.03 $\pm$   0.04 &  10.46 $\pm$   0.37 &   0.84 $\pm$   0.04 &   14.99 &   \\
       KSP-DW48 & 11:12:23.4 & -25:50:52 &  17.75 &   1.01 &   0.90 &  23.75 $\pm$   0.07 &   7.80 $\pm$   0.52 &   0.74 $\pm$   0.08 &    9.38 &   \\
       KSP-DW49 & 11:12:34.8 & -28:01:43 &  18.82 &   0.99 &   0.68 &  23.30 $\pm$   0.12 &   4.41 $\pm$   0.34 &   0.51 $\pm$   0.09 &    2.18 &   \\
       KSP-DW50 & 11:12:51.6 & -27:07:33 &  18.19 &   0.98 &   0.84 &  23.30 $\pm$   0.06 &   5.55 $\pm$   0.28 &   0.63 $\pm$   0.06 &    4.89 &   \\
       KSP-DW51 & 11:12:53.2 & -26:29:03 &  18.62 &   1.06 &   0.90 &  23.68 $\pm$   0.10 &   5.24 $\pm$   0.42 &   0.68 $\pm$   0.09 &    5.56 &   \\
       KSP-DW52 & 11:12:56.2 & -27:26:41 &  20.18 &   1.16 &   0.87 &  24.60 $\pm$   0.18 &   3.67 $\pm$   0.71 &   0.78 $\pm$   0.25 &    4.77 &   \\
       KSP-DW53 & 11:12:56.8 & -26:39:55 &  19.51 &   0.82 &   0.29 &  25.41 $\pm$   0.30 &   8.82 $\pm$   1.32 &   0.40 $\pm$   0.27 &    1.39 &   \\
       KSP-DW54 & 11:13:01.0 & -26:51:23 &  18.23 &   0.91 &   0.80 &  25.04 $\pm$   0.20 &  11.00 $\pm$   1.83 &   0.78 $\pm$   0.26 &   14.42 & U  \\
       KSP-DW55 & 11:13:09.0 & -26:19:56 &  18.52 &   1.13 &   0.64 &  23.35 $\pm$   0.28 &   3.33 $\pm$   1.13 &   1.11 $\pm$   0.30 &    5.90 &   \\
       KSP-DW56 & 11:13:27.9 & -26:56:07 &  20.15 &   0.91 &   0.67 &  24.11 $\pm$   0.12 &   3.29 $\pm$   0.36 &   0.62 $\pm$   0.16 &    2.83 &   \\
       KSP-DW57 & 11:13:32.6 & -26:32:48 &  18.29 &   1.11 &   1.09 &  24.37 $\pm$   0.12 &   8.40 $\pm$   0.79 &   0.68 $\pm$   0.14 &    8.92 &   \\
       KSP-DW58 & 11:13:36.3 & -26:33:49 &  17.85 &   0.91 &   0.71 &  23.86 $\pm$   0.08 &   8.16 $\pm$   0.57 &   0.68 $\pm$   0.07 &    8.50 & N \\
       KSP-DW59 & 11:13:48.1 & -26:07:58 &  20.38 &   1.07 &   0.75 &  25.15 $\pm$   0.17 &   5.17 $\pm$   0.66 &   0.44 $\pm$   0.22 &    1.38 &   \\
       KSP-DW60 & 11:13:55.2 & -26:22:20 &  16.98 &   0.86 &   0.61 &  23.52 $\pm$   0.06 &   8.92 $\pm$   0.54 &   0.89 $\pm$   0.06 &   13.56 & N,U \\
    ESO503-G001 & 11:14:02.2 & -26:21:56 &  13.00 &   1.11 &   0.90 &  19.08 $\pm$   0.06 &   5.87 $\pm$   0.27 &   1.11 $\pm$   0.03 &   10.43 &   \\
       KSP-DW61 & 11:14:13.1 & -26:31:38 &  19.57 &   0.98 &   1.05 &  23.86 $\pm$   0.13 &   3.47 $\pm$   0.52 &   0.77 $\pm$   0.18 &    4.47 &   \\
       KSP-DW62 & 11:14:14.4 & -26:30:47 &  16.57 &   1.21 &   0.71 &  24.74 $\pm$   0.07 &  23.93 $\pm$   1.09 &   0.52 $\pm$   0.07 &   12.45 & U  \\
       KSP-DW63 & 11:14:16.9 & -27:04:44 &  18.44 &   0.90 &   0.45 &  23.87 $\pm$   0.27 &   5.64 $\pm$   1.29 &   0.83 $\pm$   0.40 &    7.90 &   \\
       KSP-DW64 & 11:14:17.2 & -26:29:51 &  18.69 &   1.31 &   0.91 &  23.72 $\pm$   0.41 &   3.63 $\pm$   1.67 &   1.11 $\pm$   0.47 &    6.44 & N \\
       KSP-DW65 & 11:14:18.9 & -27:20:56 &  18.33 &   0.86 &   0.83 &  24.27 $\pm$   0.85 &   1.70 $\pm$   2.30 &   1.95 $\pm$   1.55 &    3.27 &   \\
       KSP-DW66 & 11:14:27.9 & -26:55:00 &  15.89 &   0.86 &   0.64 &  23.52 $\pm$   0.05 &  14.32 $\pm$   0.82 &   0.92 $\pm$   0.06 &   22.47 & U  \\
       KSP-DW67 & 11:14:29.6 & -25:16:10 &  16.32 &   1.05 &   0.77 &  22.51 $\pm$   0.04 &   7.79 $\pm$   0.27 &   0.86 $\pm$   0.03 &   11.43 &   \\
       KSP-DW68 & 11:14:36.4 & -26:51:15 &  18.70 &   0.94 &   0.70 &  23.71 $\pm$   0.16 &   4.98 $\pm$   0.65 &   0.73 $\pm$   0.16 &    5.85 &   \\
       KSP-DW69 & 11:14:54.3 & -27:49:00 &  19.26 &   0.95 &   0.85 &  22.98 $\pm$   0.09 &   2.71 $\pm$   0.25 &   0.75 $\pm$   0.10 &    3.34 &   \\
       KSP-DW70 & 11:15:24.1 & -26:45:14 &  17.23 &   1.01 &   0.82 &  22.96 $\pm$   0.04 &   6.95 $\pm$   0.28 &   0.73 $\pm$   0.04 &    8.21 &   \\
       KSP-DW71 & 11:15:58.1 & -27:29:57 &  19.76 &   0.68 &   0.25 &  23.65 $\pm$   0.08 &   3.24 $\pm$   0.25 &   0.58 $\pm$   0.12 &    2.35 &   \\
       KSP-DW72 & 11:16:21.2 & -26:24:26 &  19.56 &   0.75 &   0.47 &  23.96 $\pm$   0.21 &   4.10 $\pm$   0.65 &   0.58 $\pm$   0.18 &    3.04 &   \\
       KSP-DW73 & 11:16:47.3 & -27:40:17 &  16.39 &   1.14 &   0.78 &  22.86 $\pm$   0.03 &   9.52 $\pm$   0.27 &   0.76 $\pm$   0.03 &   12.07 &   \\
       KSP-DW74 & 11:17:55.0 & -26:07:35 &  18.37 &   1.01 &   0.75 &  23.03 $\pm$   0.75 &   1.99 $\pm$   2.03 &   1.49 $\pm$   0.82 &    3.83 & N \\
       KSP-DW75 & 11:18:26.3 & -26:41:52 &  18.93 &   0.97 &   0.74 &  23.29 $\pm$   0.20 &   3.21 $\pm$   0.62 &   0.91 $\pm$   0.16 &    4.97 &   \\
       KSP-DW76 & 11:18:26.4 & -26:11:39 &  16.46 &   0.84 &   0.60 &  22.83 $\pm$   0.03 &   9.02 $\pm$   0.30 &   0.77 $\pm$   0.03 &   11.62 &   \\
       KSP-DW77 & 11:18:37.1 & -26:55:09 &  16.53 &   0.55 &   0.37 &  23.14 $\pm$   0.05 &   9.81 $\pm$   0.57 &   0.81 $\pm$   0.06 &   13.45 &   \\
\enddata
\tablenotetext{~}{
$^a$ KSP-DW** candidates are newly discovered objects in KSP, 
$^b$ $I$ is $I$-band total magnitude derived from S{\'e}rsic fit.
$^c$ $(V-I)$ and $(B-V)$ are colors without the extinction correction. 
$^d$ $\mu_{0,I}$, $r_{0,I}$, $n_{I}$, and $r_{e,I}$ are central surface brightness, scale length, curvature index, and effective radius derived for the $I$-band  S{\'e}rsic fits, respectively.
$^e$ 'N' and 'U' flag a nucleated galaxy candidate and UDG candidate, respectively. 
}
\end{deluxetable}
\clearpage

\clearpage

\clearpage

\begin{figure}
\epsscale{1.0} 
\plotone{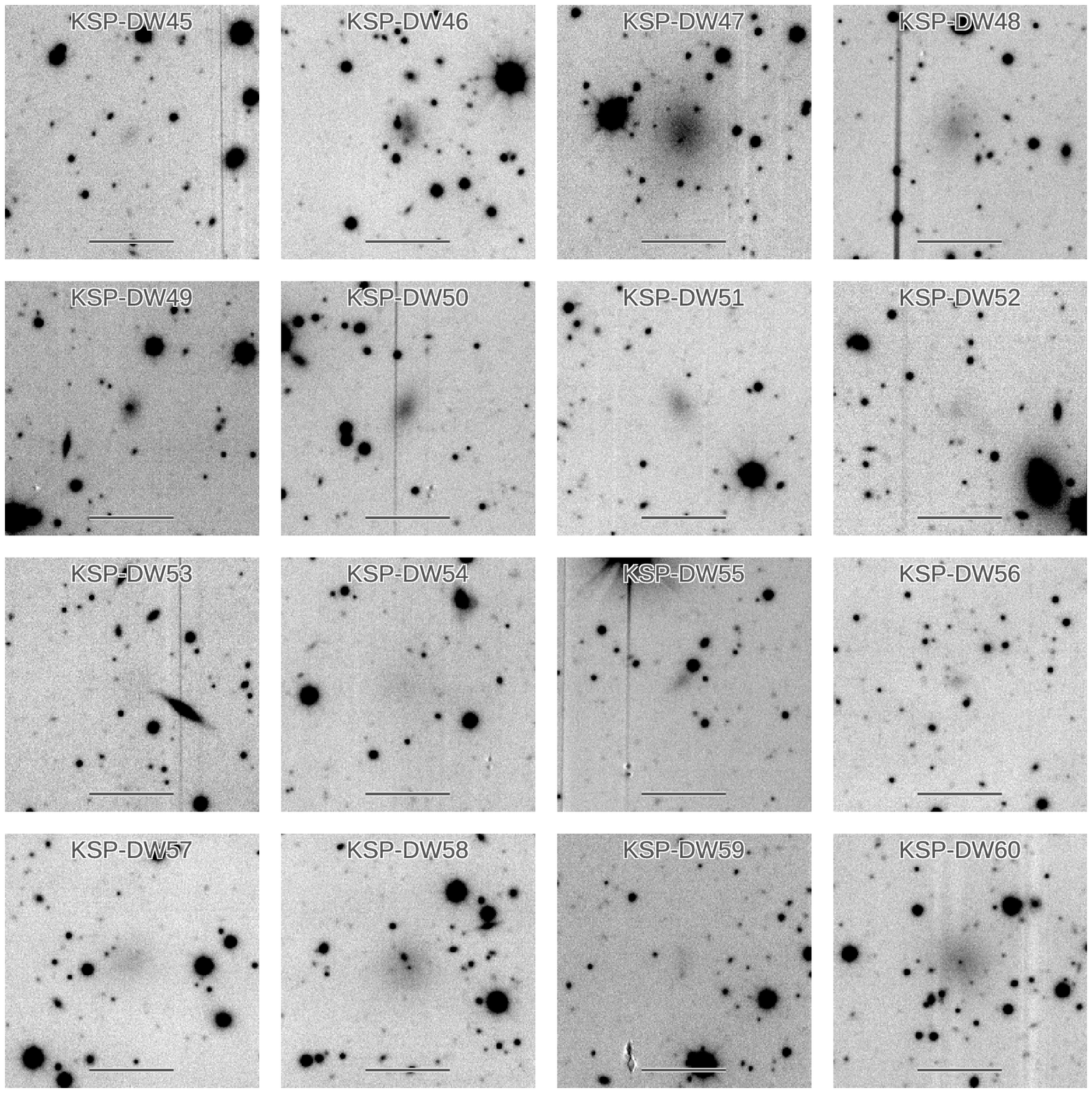}
\centerline{Fig. 2. --- {\it Continued}}
\end{figure}
\clearpage

\begin{figure}
\epsscale{1.0} 
\plotone{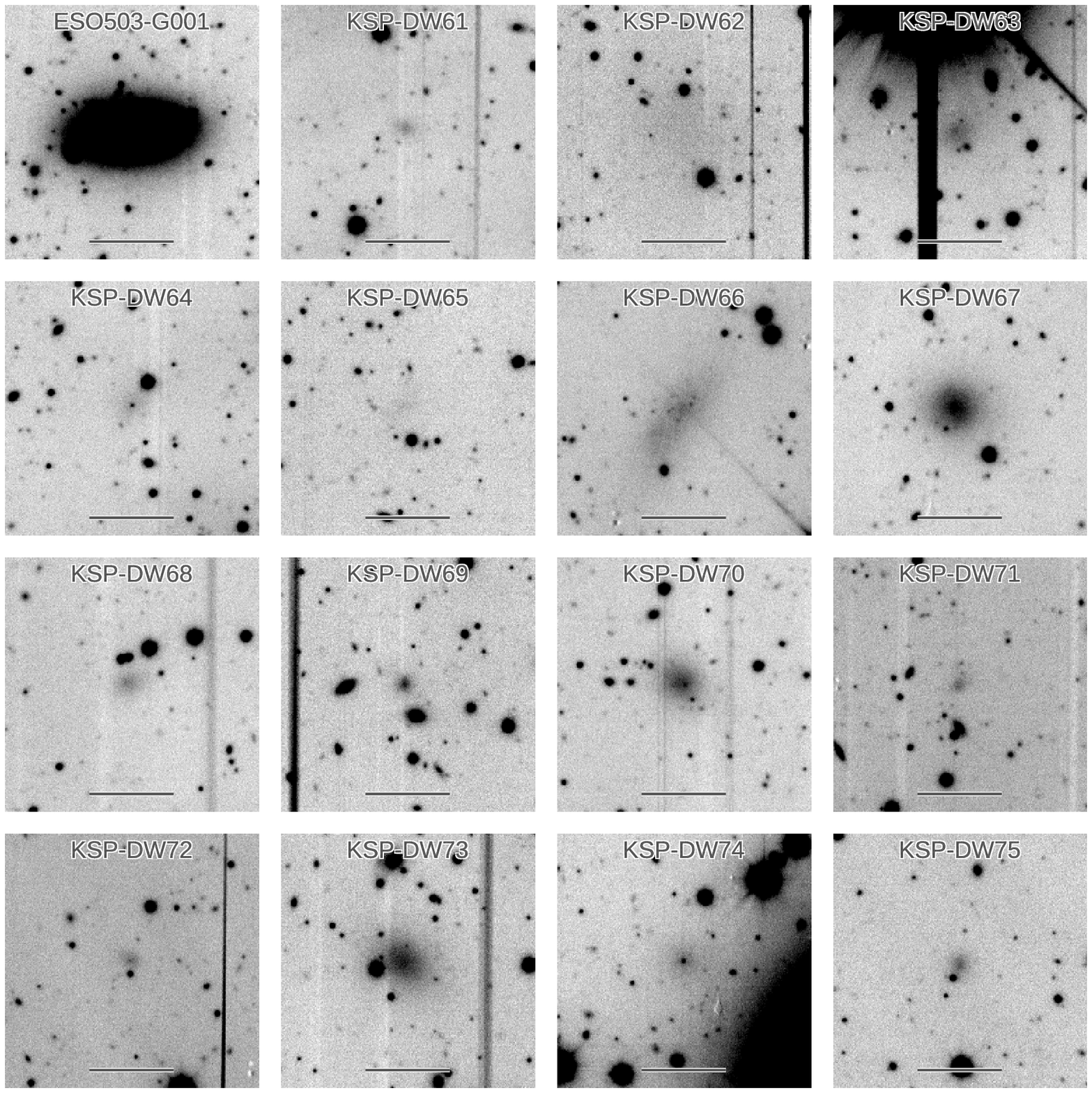}
\centerline{Fig. 2. --- {\it Continued}}
\end{figure}
\clearpage

\begin{figure}
\epsscale{1.0} 
\plotone{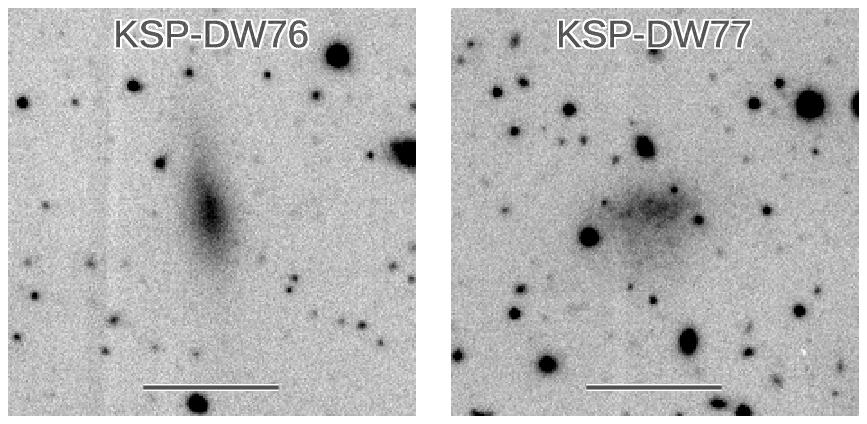}
\centerline{Fig. 2. --- {\it Continued}}
\end{figure}
\clearpage

\begin{figure}
\epsscale{1.0}
\plotone{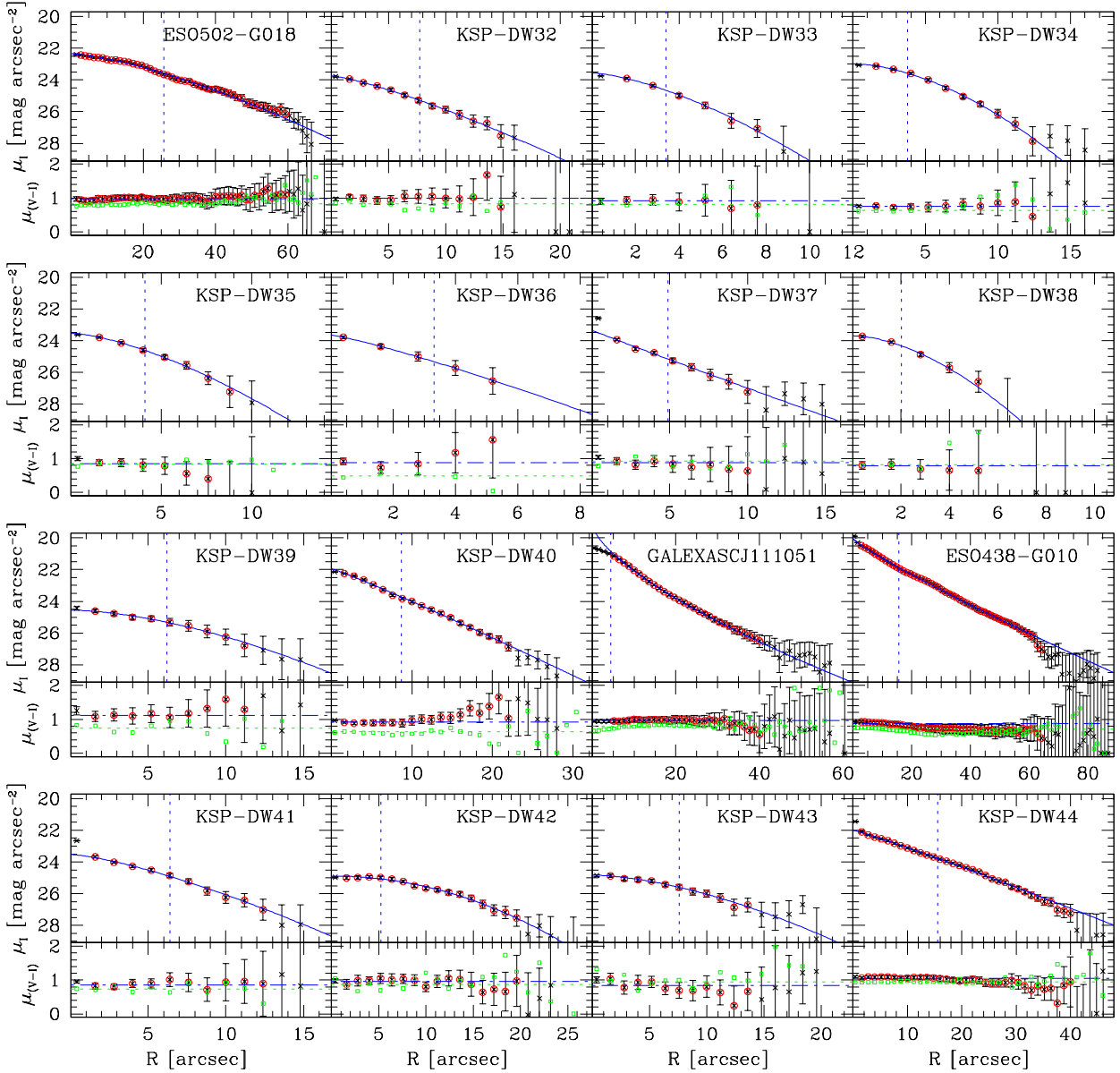}
\caption{ 
Surface brightness and color profiles of the dwarf galaxy candidates. 
The red open circles in each panel indicate surface brightnesses and $(V-I)$ colors 
  for which the S{\'e}rsic fit is shown by the solid lines.
The green open squares are $(B-V)$ colors and
  the vertical dotted lines represent the effective radii. 
The mean colors of $(V-I)$ and $(B-V)$ are presented by blue dot-dashed and
 green dotted lines, respectively.
\label{fig-radimu0a}}
\end{figure}
\clearpage

\begin{figure}
\epsscale{1.0}
\plotone{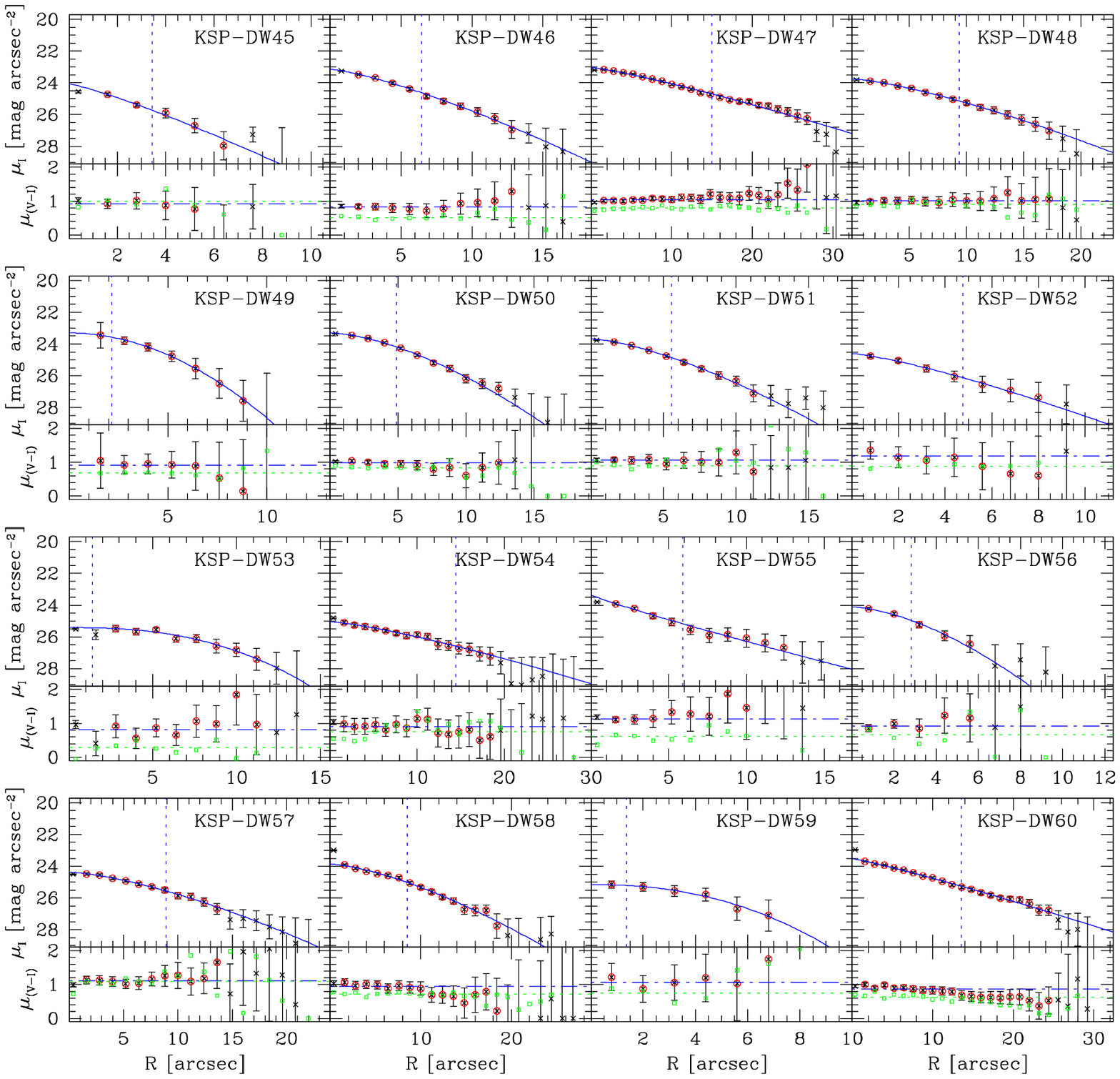}
\centerline{Fig. 3. --- {\it Continued}}
\end{figure}
\clearpage

\begin{figure}
\epsscale{1.0}
\plotone{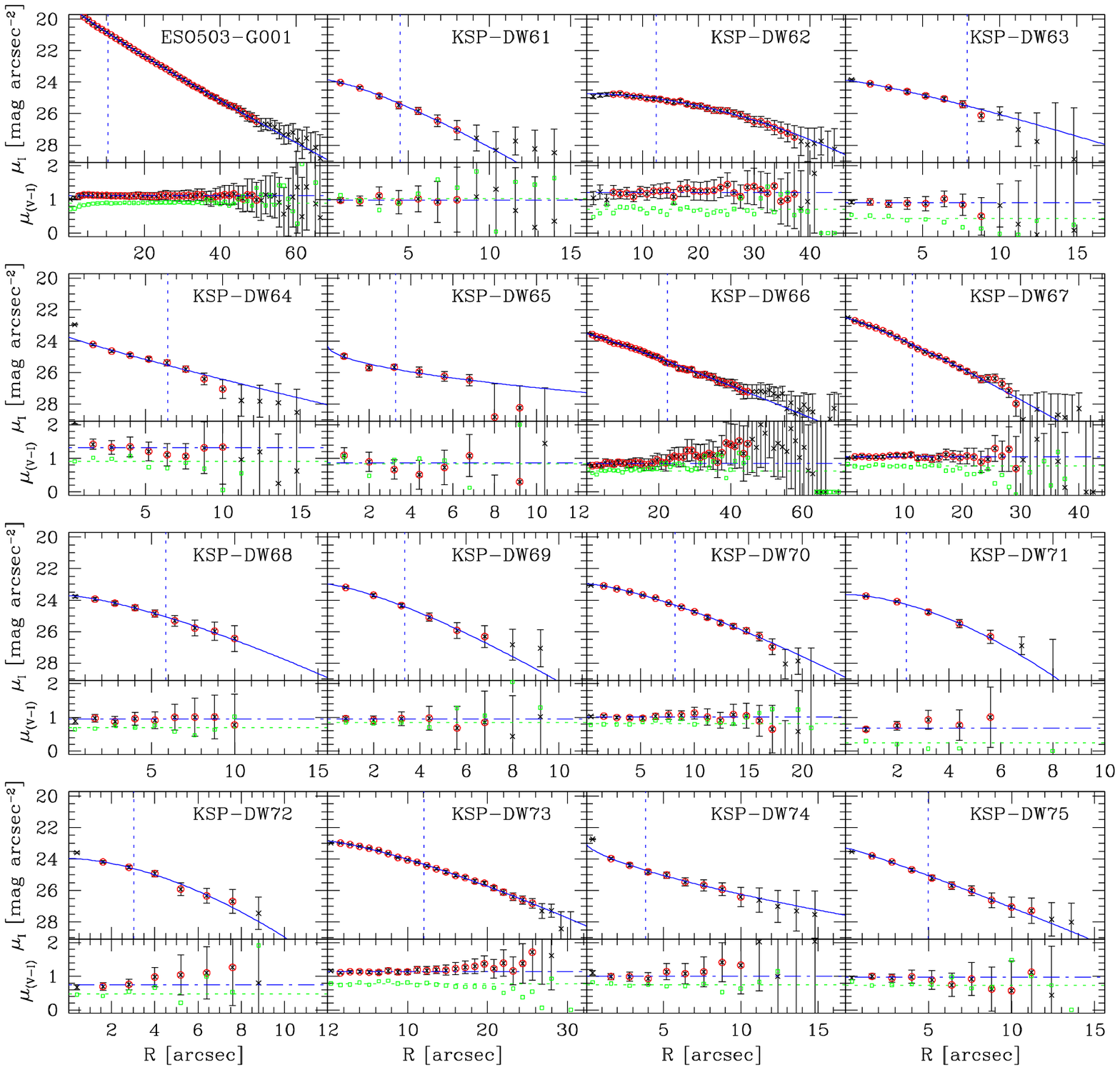}
\centerline{Fig. 3. --- {\it Continued}}
\end{figure}
\clearpage

\begin{figure}
\epsscale{0.5}
\plotone{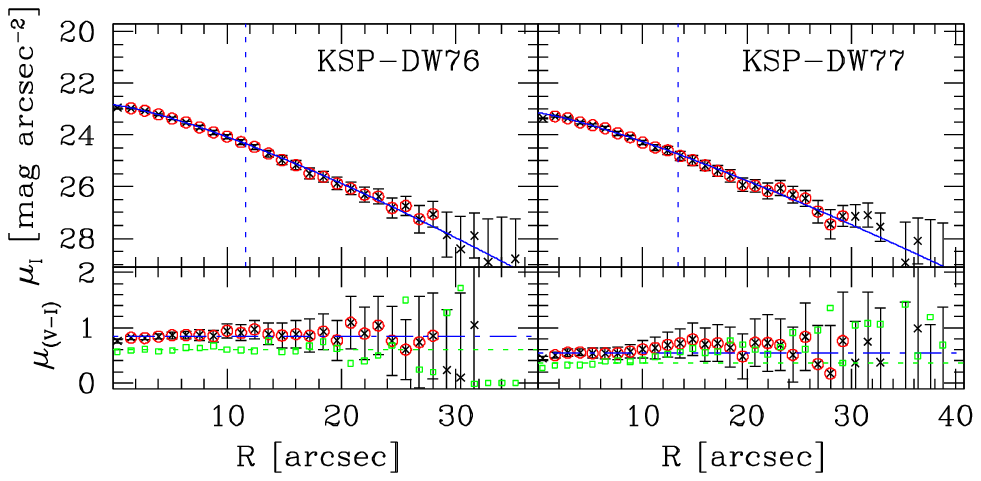}
\centerline{Fig. 3. --- {\it Continued}}
\end{figure}
\clearpage

\begin{figure}
\epsscale{0.9} 
\plotone{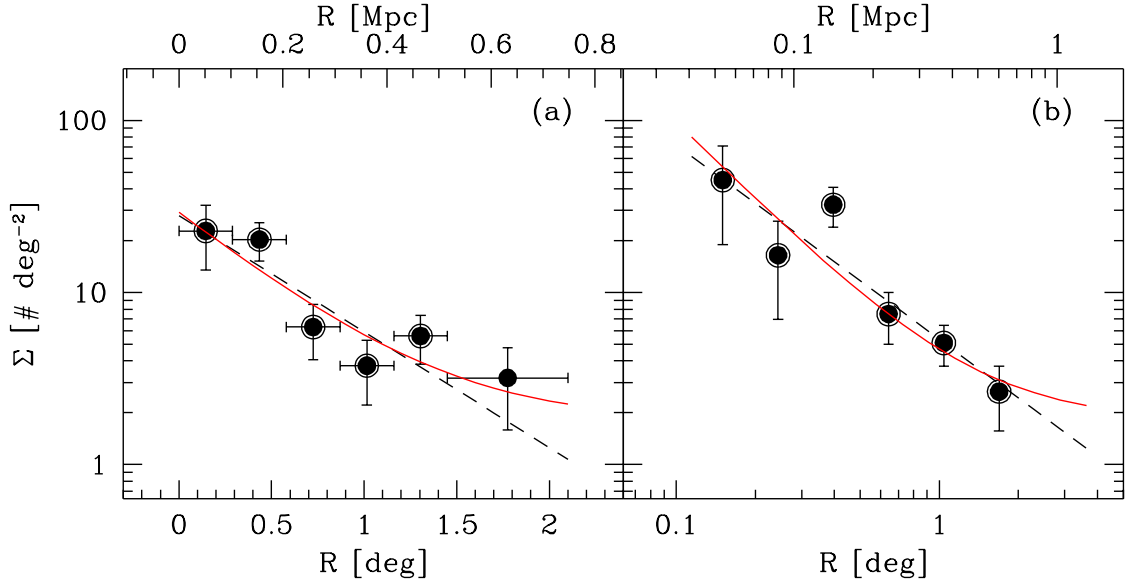}
\caption{Radial, projected number density of the dwarf galaxy candidates in the NGC 3585 field.
(a) The radial binning is done in equal linear intervals except for the last bin. 
The dashed line and the solid curve represent the best-fit exponential function 
  for $R\lesssim1.5$ deg and the exponential + $constant$ function, respectively.
(b) The radial binning is done in equal logarithmic intervals, except for the last bin.
The \textcolor{black}{black} dashed line and the \textcolor{black}{red} solid curve represent the best-fit power-law functions without and with an added $constant$ value, respectively.
\label{fig-rnden}}
\end{figure}
\clearpage

\begin{figure}
\epsscale{0.6}
\plotone{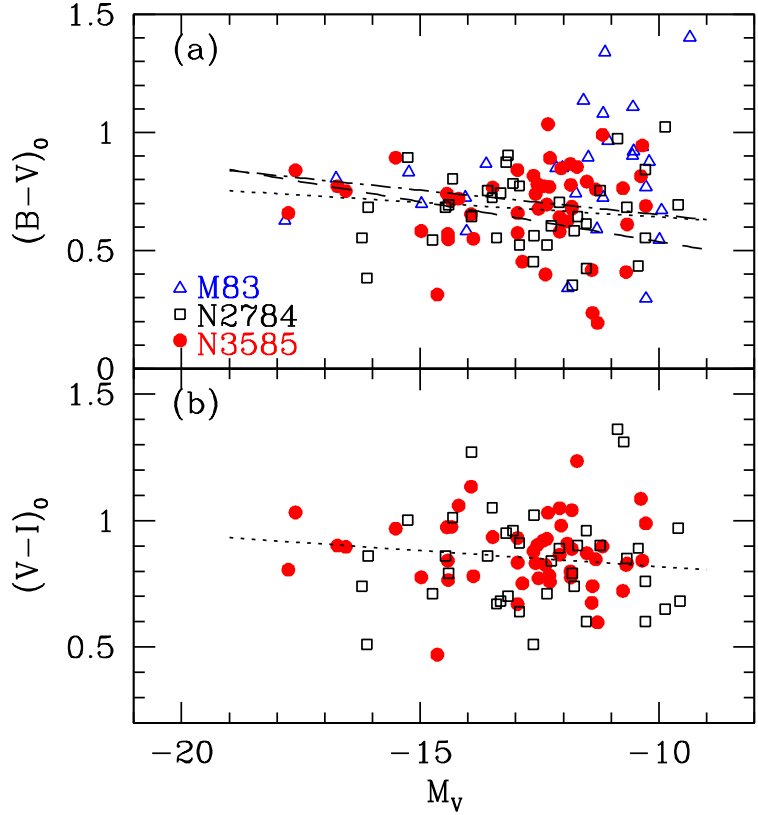}
\caption{ 
Color-magnitude diagrams of the dwarf galaxy candidates in NGC 3585 fields.
The circles, squares, and triangles represent the dwarf galaxy candidates 
  in the NGC 3585 group (this study), NGC 2784 group \citep{par17}, 
  and the M83 group \citep{mul15}, respectively. 
The dotted lines indicate the color-magnitude relations obtained 
  from the dwarf galaxy candidates in the NGC 2784 group and the NGC 3585 group, while
  the dot-dashed and the dashed lines are
  obtained from the early-type galaxies in
  the Virgo cluster \citep{lis08} and the Ursa Major cluster \citep{pak14}, respectively.
\label{fig-cmd}}
\end{figure}
\clearpage

\begin{figure}
\epsscale{0.6}
\plotone{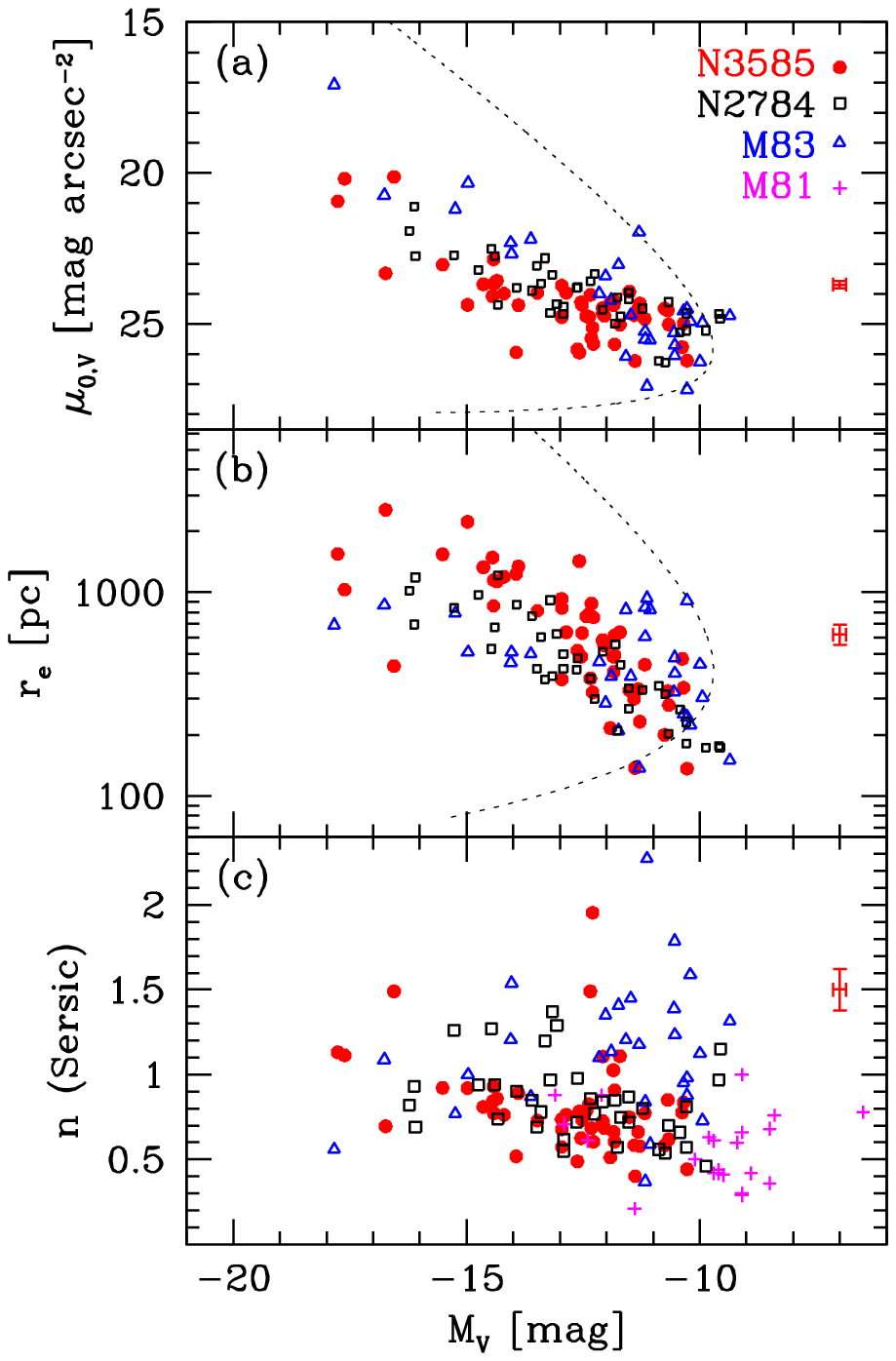}
\caption{ 
Central surface brightness (a), effective radius (b), and S{\'e}rsic-$n$ (c) versus total absolute magnitude for the dwarf galaxy candidates.
The circles, squares, triangles, and pluses indicate the dwarf galaxy candidates 
in the NGC 3585 group (this study), NGC 2784 group \citep{par17}, M83 group \citep{mul15},  
 and  M81 group \citep{chi09},
respectively. 
\textcolor{black}{
  The dotted curves represent the completeness limits for our survey (see the Section \ref{dwsearch}).
  The error bar in each panel represents the mean value of uncertainties in each parameter.}
\label{fig-censurface}}
\end{figure}
\clearpage

\begin{figure}
\epsscale{0.6}
\plotone{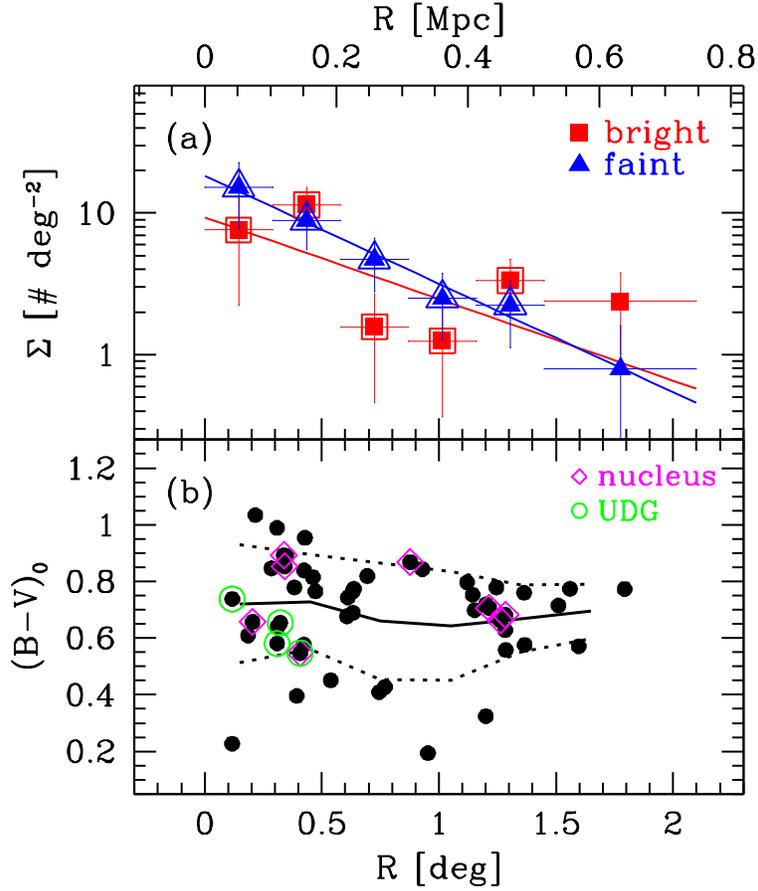}
\caption{ 
Radial number density profile (a) and color distribution (b) 
  for the dwarf galaxy candidates in the NGC 3585 group.
The squares and triangles in (a) are the number densities
  for the bright ($M_V<-12.5$) and faint ($M_V\geq-12.5$) dwarf samples, respectively.
The solid lines represent exponential fits to each sample.
Only the measurements represented by larger, open symbols are used 
in the fitting.
The solid and dotted curves in (b) indicate
  the mean color value and standard deviation in each radial bin, respectively.
\textcolor{black}{The green open circles and magenta open diamonds represent the UDG and nucleated dwarf galaxy candidates in the NGC 3585 group. respectively.}
\label{fig-radcoldenudg}}
\end{figure}
\clearpage

\begin{figure}
\epsscale{0.9}
\plotone{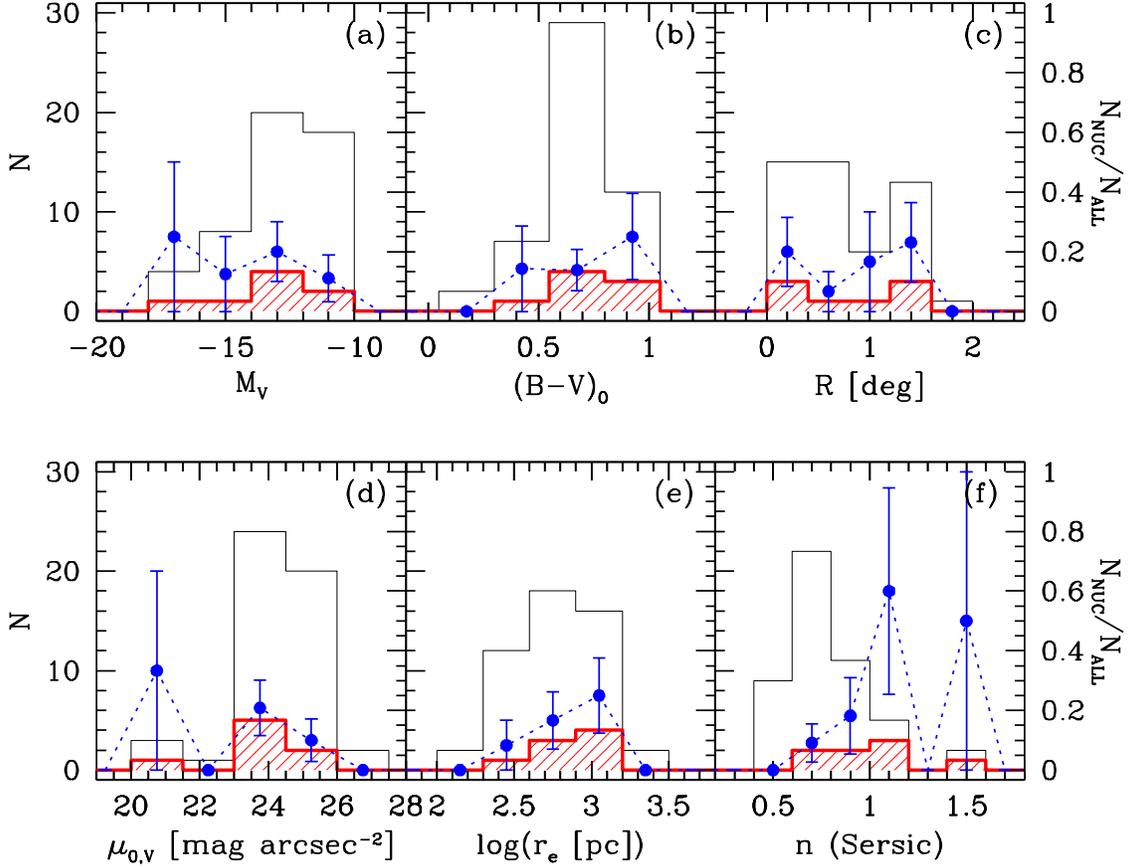}
\caption{ 
Number and number ratio distributions of nucleated dwarf galaxy candidates in terms of luminosity (a), color (b), 
 \textcolor{black}{projected distance from the group center (c),} 
  central surface brightness (d), log effective radius (e), and S{\'e}rsic-$n$ (f).
The solid and hatched histograms represent all the dwarf galaxy candidates and the nucleated dwarf galaxy candidates 
  in the NGC 3585 group, respectively.
The dotted curves with the filled circles indicate the fraction of nucleated dwarf galaxy candidates ($N_{NUC}/N_{ALL}$)
\textcolor{black}{and error bars represent Poisson uncertainties.} 
\label{fig-nuc}}
\end{figure}
\clearpage

\begin{figure}
\epsscale{1.0}
\plotone{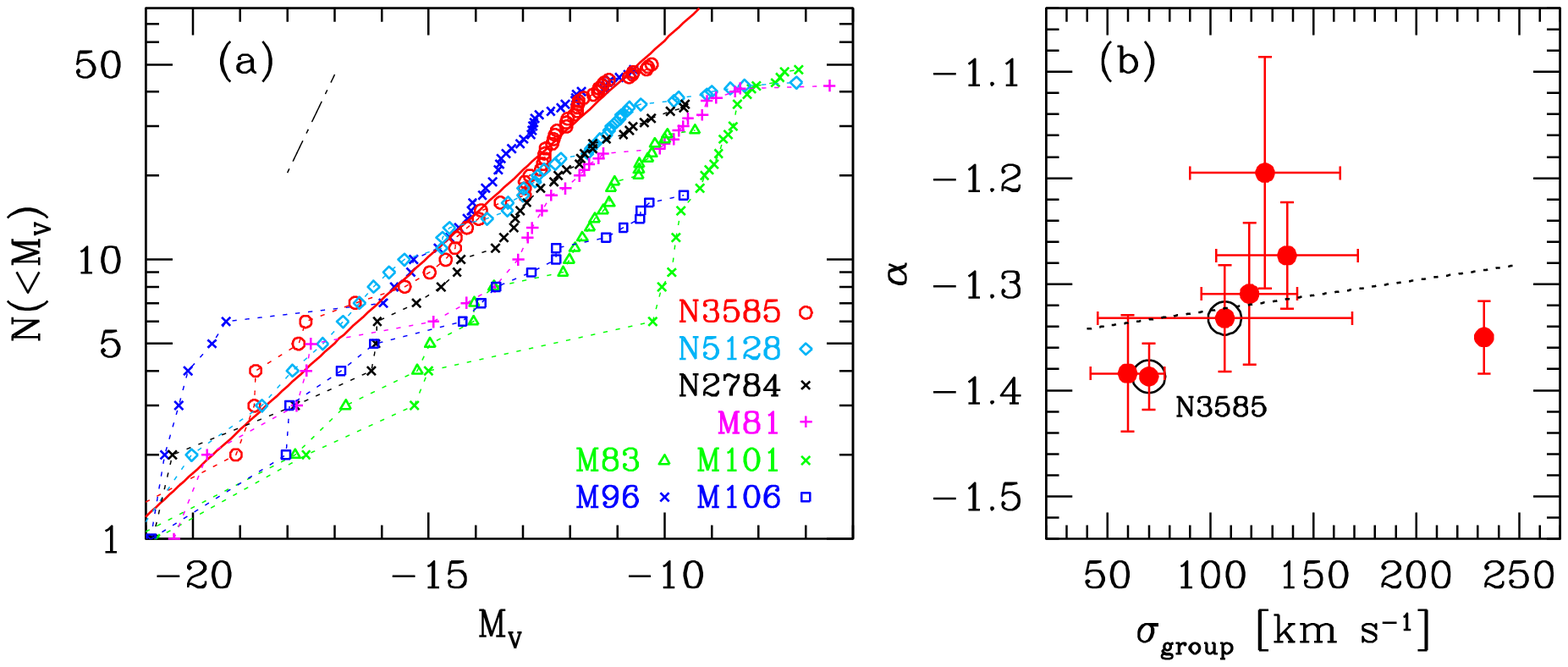}
\caption{
(a) Cumulative LF of galaxies in the NGC 3585 group and several other galaxy groups. 
\textcolor{black}{The red circles, cyan diamonds, black crosses, magenta pluses, green triangles, 
blue crosses, green crosses, 
and blue squares}
represent the galaxies 
in NGC 3585 group (this study), 
  NGC 5128 group \citep{tul15,crn16}, 
  NGC 2784 group \citep{par17}, 
  M81 group \citep{chi09}, 
  M83 group \citep{mul15},
  \textcolor{black}{M96 group \citep{mul18}, M101 group \citep{ben17,ben19},} 
  and M106  group \citep{kim11}, respectively.
The solid line indicates the best fitting cumulative Schechter function of the dwarf galaxy candidates in the NGC 3585 group. 
The dot-dashed line represents the cumulative Schechter function form for the faint-end slope ($\alpha=-1.8$) of halo masses expected from the $\Lambda$CDM model \citep{tre02}.
(b) LF slopes ($\alpha$) of the groups 
as a function of 
the velocity dispersion ($\sigma_{group}$) of the member galaxies in each group.
The measurement for the NGC 3585 group is labeled and the other open symbol represents the result for NGC 2784 \citep{par17}. The
  other filled circles represent other groups obtained from the literature data.
The dotted line represents the linear least-square fit to the data.
\label{fig-clfalpha}}
\end{figure}
\clearpage

\end{document}